\documentclass[fleqn,usenatbib]{mnras}

\usepackage{newtxtext,newtxmath}

\usepackage[T1]{fontenc}
\usepackage{ae,aecompl}

\DeclareRobustCommand{\VAN}[3]{#2}
\let\VANthebibliography\thebibliography
\def\thebibliography{\DeclareRobustCommand{\VAN}[3]{##3}\VANthebibliography}

\usepackage{graphicx}	
\usepackage{amsmath}	
\usepackage{footnote}
\usepackage{tablefootnote}

\newcommand{\src}{4U 1636--536 }

\title[Broadband AstroSat study of \src bursts]{Broadband time-resolved spectroscopy of thermonuclear X-ray bursts from 4U 1636-536 using AstroSat}

\author[U. Kashyap et al.]{
Unnati Kashyap $^{1}$\thanks{E-mail: phd1801121005@iiti.ac.in},
Biki Ram $^{1}$,
Tolga G\"uver $^{2, 3}$,
Manoneeta Chakraborty$^{1}$
\\
$^{1}$ Department of Astronomy, Astrophysics and Space Engineering (DAASE),
Indian Institute of Technology Indore, Khandwa Road, Simrol, Indore 453552, India \\
$^{2}$ Istanbul University, Science Faculty, Department of Astronomy and Space Sciences, Beyaz\i t, 34119, \.Istanbul, Turkey\\
$^{3}$ Istanbul University Observatory Research and Application Center, Istanbul University 34119, \.Istanbul Turkey\\ 
}

\date{Accepted XXX. Received YYY; in original form ZZZ}

\pubyear{2021}

\begin{document}
\label{firstpage}
\pagerange{\pageref{firstpage}--\pageref{lastpage}}

 \maketitle
\begin{abstract}
Broadband spectral studies of Type-I X-ray bursts can put strong constraints on the physics of burst spectra as well as their interaction with the environment. We present the results obtained from the broadband time-resolved spectroscopy of 15 thermonuclear bursts detected simultaneously from the neutron star atoll source 4U 1636--536  using LAXPC and SXT onboard AstroSat. During the observations with AstroSat, the Low mass X-ray binary (LMXB) 4U 1636--536 is observed to show a modest spectral evolution within the island state. The broadband burst spectra are observed to show an excess in addition to the thermal emission from the neutron star surface near the peak of the bursts. We investigate the interpretation of the excess observed near the peak of the burst as re-emission/reprocessing of the photons by the accretion disk/corona or scattering of the photons in the neutron star atmosphere or the enhanced persistent emission due to Poynting–Robertson drag. This is the first reported broadband simultaneous study of Type-I bursts using LAXPC and SXT onboard AstroSat. This kind of study may provide a better understanding of the burst-accretion interaction and how the bursts influence the overall accretion process contributed by the accretion disk as well as the corona. 
\end{abstract}



\section{Introduction}



Thermonuclear X-ray bursts (Type-I) occur due to the unstable burning of primarily hydrogen and/or helium at the surface of the accreting neutron star in Low Mass X-ray Binaries (LMXBs). During these thermonuclear ignition events, the source flux increases by several times than the persistent level \citep{2008ApJS..179..360G, 2009MNRAS.398..368Z,1976Natur.263..101W}. These thermonuclear bursts are characterized by fast rise times ranging between 1 and 10 seconds, long decay times ranging from ten to hundreds of seconds, and recurrence times of hours to days. Following the peak, the burst undergoes an exponential decay \citep{1993SSRv...62..223L}. Several meters thick layers are formed via the accumulation of matter on the surface of the neutron star from the binary companion \citep{2008ApJS..179..360G}. The burst properties strongly depend on the mass accretion rate \citep{1981ApJ...247..267F, 2011MNRAS.413.1913Z,2021ASSL..461..209G} in addition to several other factors such as accreted fuel, the spin of the neutron star, compactness, metallicity as well as the thermal state of the accreting neutron star \citep{2003ApJ...599..419N,1995ApJ...438..852B,1998ASIC..515..419B,2004ApJ...603L..37C}. The physics behind the occurrence of Type I X-ray burst is very complex, but the burst spectra were usually well described by the simple blackbody function \citep{2001ApJ...554L..59W,2012ApJ...747...76G}. The time-resolved spectroscopic analysis of thermonuclear bursts shows that the burst spectra are described by evolving Planckian functions with temperature ranging from 1-3 keV  \citep{2006csxs.book..113S, 2018SSRv..214...15D,2021ASSL..461..209G}. Deviations from the Planckian model were found during the RXTE/PCA observations indicating an excess in the burst spectra \citep{2013ApJ...772...94W,2015ApJ...801...60W}. This kind of excess may be produced by reflections of burst radiation from the disk or reprocessing of the burst photons from the corona or enhancement of the accretion flow due to the Poynting–Robertson radiation drag on the disk \citep{2013A&A...553A..83I, 2013ApJ...767L..37D, 2017ApJ...836..111K,  2018ApJ...855L...4K, 2013ApJ...772...94W,2021MNRAS.tmp.2417R}. The detection of the iron lines during superbursts reveals the reprocessing/reflection of the burst emission by the inner accretion disk and impacts of the bursts on the accretion disk \citep{2004MNRAS.351...57B,2004ApJ...602L.105B,2014ApJ...797L..23K}.

The LMXB neutron star atoll source 4U 1636-536 is a very frequent and prolific X-ray burster.  A total of 664 unique bursts were detected from this source starting from 1996 February 8 through 2012 May 3 with a mean burst rate of 0.26 $hr^{-1}$ \citep{2020ApJS..249...32G}. Most bursts from this source are detected and studied with instruments that are sensitive to relatively high photon energies and the corresponding spectra were usually fitted with a standard Planckian spectrum \citep{2008ApJS..179..360G,2014ApJ...797L..23K,2019MNRAS.482.4397B}.

In this work, we analyze a total of 15 Type I X-ray bursts detected simultaneously by LAXPC and SXT onboard AstroSat. We study and investigate the interpretations of the broadband spectra of these bursts. We present the results obtained from the broadband (LAXPC+SXT) time-resolved spectroscopy of all the bursts detected simultaneously. This kind of joint broadband study of thermonuclear bursts provides a better opportunity to study burst-accretion interactions and the impact of bursts on the accretion environment. We also present a detailed analysis of the excess detected in the burst spectra. Through this work, we show that the broadband capability of AstroSat enables an insightful study of the deviations from the Planckian spectrum during a thermonuclear burst and their physical origin. 

\section{Observations and Data Reduction}

\begin{table}
\centering
\caption{ Observation details of the bursts detected in the LAXPC data of \src
\label{laxpcbursts}}
\begin{tabular}{|c|c|c|}
\hline
Observation ID && No. of Bursts detected \\ 
\hline
9000001326 &&3 \\
9000001574 &&6\\
9000002084 && 2 \\
9000002278 &&2\\
9000002316 && 14\\
\hline
\end{tabular}
\end{table}

\begin{table*}
\centering
\caption{Details of the  bursts detected simultaneously with SXT as well as LAXPC onboard AstroSat from 4U 1636-536 
\label{burstsample}}
\begin{tabular}{ccccccccccccccccc}
\hline
Burst ID & Observation ID & Date (dd-mm-yyyy) & Start time Burst (MET/UTC) &  &  & Duration (s) & Peak Flux$^a$ & Fluence$^b$  \\
\hline

B1& 9000001326 & 21-06-2017 & 235769000.50 / 19:23:19.50 &&   &42.79 &$3.80_{-0.09}^{+0.09}$&$6.09^{+1.78}_{-1.78}$\\
B2& & & 235780295.43 / 22:31:34.43 &&& 57.05& $17.38_{-0.00}^{+0.40}$&$24.56^{+7.55}_{-7.55}$\\
\hline
B3 & 9000001574 & 02-10-2017 & 244654733.96 / 15:38:52.96 &&& 47.55 & $7.41_{-0.17}^{+0.17}$&$8.96^{ +3.45}_{-3.45}$\\
\hline
B4 & 9000001574 & 03-10-2017 & 244707001.09 / 06:10:00.09 &&& 23.77&$2.09_{ -0.09}^{+0.10}$&$2.42^{+0.67}_{-0.67}$\\
\hline
B5 & 9000002084 & 09-05-2018 &  263585754.65 / 18:15:53.65 &&& 23.77 &$4.57_{-0.10}^{+0.11}$& $4.08_{ -1.70}^{+1.70}$\\
\hline
B6 & 9000002278 & 06-08-2018 & 271268546.74 / 16:22:25.74 &&& 76.07 &$14.79_{-0.34}^{+0.34}$&$29.54_{  -8.76}^{+8.76}$\\ 
B7&&& 271278876.47 / 19:14:35.47 &&& 66.57 &$20.89_{0.00}^{ 0.49}$&$32.55_{-11.37}^{+11.37}$\\ 
\hline
B8 & 9000002316 & 18-08-2018 & 272284280.83 / 10:31:19.83 &&& 33.28 &$5.75_{-0.13}^{+0.13}$&$7.70_{ -2.40}^{+2.40}$\\
\hline
B9 & 9000002316 & 19-08-2018 & 272389159.87 / 15:39:18.87 &&& 47.55 & $14.45_{-0.33}^{+0.00}$&$21.75_{  -6.11}^{+6.11}$\\
B10 &&& 272390686.29 / 16:04:45.29 &&& 38.04 &$5.50_{-0.13}^{+0.13}$&$7.79_{-2.54}^{+2.54}$\\

B11 &&& 272401025.50 / 18:57:04.50 &&& 71.32 &$13.18_{-0.30}^{+0.00}$&$20.50_{-6.75}^{+6.75}$\\
\hline
B12 & 9000002316 & 20-08-2018 & 272419666.72 / 00:07:45.72 &&& 71.32 & $12.88_{-0.29}^{+0.30}$& $20.65_{-6.50}^{+6.50}$\\
B13 &&& 272448343.01 / 08:05:42.01 &&& 52.30 &$13.80_{- 0.00}^{+0.32}$&$27.15_{-7.02}^{+7.02}$\\
B14 &&& 272478238.91 / 16:23:57.91 &&& 71.32&$16.98_{ -0.00}^{+0.40}$& $26.90_{-8.53}^{+8.53}$\\
B15 &&& 272488604.11 / 19:16:43.11 &&& 57.05 &$16.98_{- 0.00}^{+0.40}$& $25.26_{-8.11}^{+8.11}$\\
\hline
\end{tabular}
\begin{flushleft}
\begin{footnotesize}
$^a$ In units of $10^{-9}$ erg/s/cm$^{2}$ \\
$^b$ In units of $10^{-8}$ erg/cm$^{2}$
\end{footnotesize}
\end{flushleft}
\end{table*}

\subsection{LAXPC}
 The LAXPC (Large Area X-ray Proportional Counter) instrument is one of the major payloads on AstroSat consisting of three identical proportional counter detector units named LAXPC10, LAXPC20, and LAXPC30. It has three identical units with the largest effective area in the mid-X-ray range 3-80 keV and a large area collection of 6000 $cm^{2}$ at 15 keV. The deadtime and time resolution of these detectors are about 42 $\mu$s and 10 $\mu$s. The good time resolution makes LAXPC ideal for fast timing analysis and time-resolved spectroscopy of thermonuclear bursts \citep{2017CSci..113..591Y}.

\subsection{SXT}
The Soft X-ray Telescope (SXT) onboard AstroSat provides spectra in the energy range of 0.3-8 keV. The effective area of SXT is 90  $cm^{2}$  at 1.5 keV, and it has a focal length of 2 meters. Its field of view is $40\:'$ and the size of the point spread function (PSF) is 3-4 arcmin. The time resolution of SXT is 2.37 s. The spatial as well as spectral resolution and the good sensitivity makes SXT favorable for spectral analysis as well as variability observations in the soft X-ray regime \citep{2017JApA...38...29S}.  

\subsection{Data}
We searched the AstroSat observation database of 4U 1636--536 starting from 21 June 2017 till 20 Aug 2018 for bursts that were simultaneously covered by LAXPC and SXT. LAXPC and SXT jointly observed this source over 8 observational epochs during this entire period (Table~\ref{burstsample}).

\subsubsection{LAXPC}

For spectro-temporal analysis using LAXPC data, event analysis (EA) mode data were used. In the event mode, the arrival times, as well as energies of each detected photons, are recorded. The details of the bursts detected by LAXPC are listed in Table~\ref{laxpcbursts}. The LAXPC data were analyzed using {\tt LaxpcSoft} software \citep{2017ApJS..231...10A}. The light curves, spectra as well as background, and response files were processed using {\tt LaxpcSoft} software. Due to gas leakage, LAXPC30 has low gain and consequently, the LAXPC 30 data were not useful. For most of the LAXPC analysis, LAXPC20 data were used as it provides the maximum gain and sensitivity.

\subsubsection{SXT}
Similar to LAXPC, we collected SXT level2 data for the same observations. In order to extract the image, light curves, as well as spectrum, these cleaned level2 event files were used. For the purpose of extracting images, light curves, as well as spectrum ftool XSELECT distributed as a part of Heasoft 6.24 package, was used. The {\tt sxtARFmodule} was used to produce appropriate ancillary response files. During spectral analysis the response file (sxt\_pc\_mat\_g0to12.rmf)  and  blank  sky  background spectrum file (SkyBkg\_comb\_EL3p5\_Cl\_Rd16p0\_v01.pha) provided by the SXT team were used.  All the data were in PC (photon counting) mode. In PC mode, data from the entire CCD ($\approx$ 36000 pixels) above specified threshold energy are collected. The time resolution of SXT in this mode is 2.37 s. The region-filtered event files were obtained using a circular region of 16 arcmins around the source location using {\tt ds9}. 

\section{Results}

\subsection{The HID}

\begin{figure}
    \centering
     \includegraphics[width=0.50\textwidth]{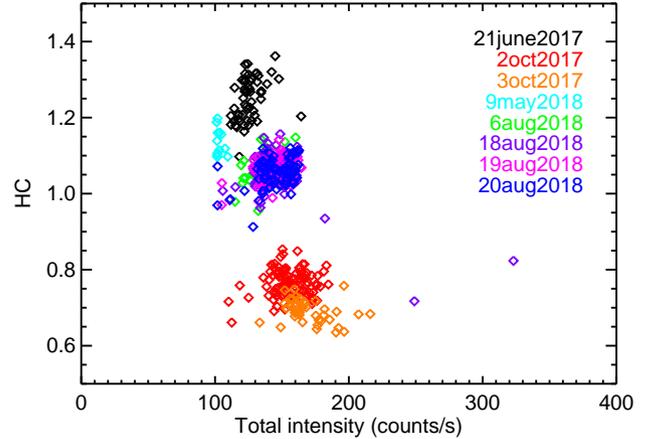}
     \caption{ The hardness-intensity diagram of the LAXPC observations of \src which contain the bursts in our sample. Different colors represent the observations on different dates.  Hard color is defined as the 16.0-9.7 keV/9.7-6.4 keV background subtracted count rate ratio and intensity as the whole energy band background subtracted count rate.  Each point corresponds to a 256 s bin size. \label{HID}}
\end{figure}

 Figure~\ref{HID} shows the Hardness Intensity diagram (HID) of the source. The hard color (HC) is defined as the ratio of the background-subtracted count rates in two energy bands 16.0-9.7 keV and 6.4-9.7 keV. The different colors demonstrate the observations on different days.  Starting from 21 June 2017, the source moves to a relatively softer state during 2 October 2017 to 3 October 2017 observations. The source is observed to be in a relatively harder state during 18 August 2018 to 20 August 2018 observations.

\subsection{Burst light curves}
\begin{figure}
    \centering
     \includegraphics[width=0.49\textwidth]{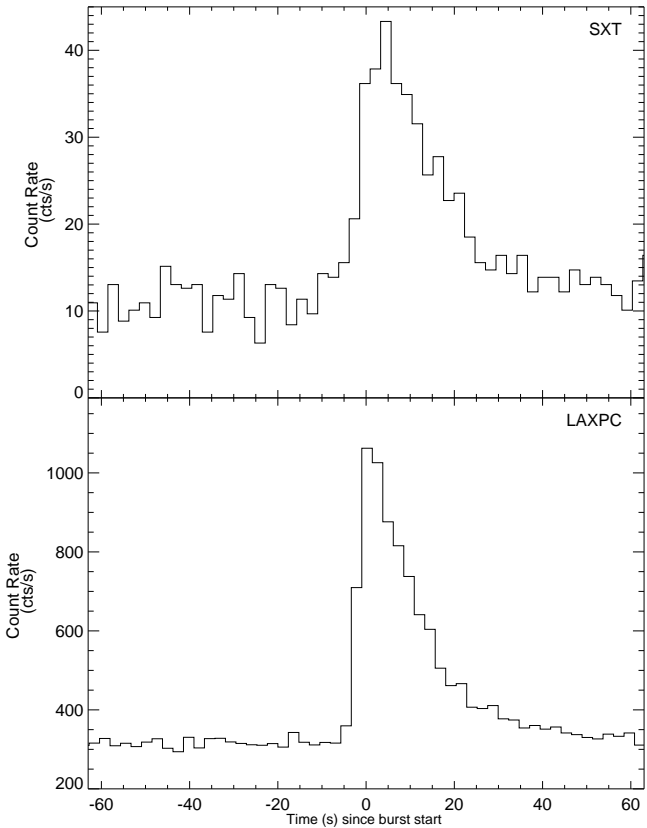}
     \caption{SXT (upper panel) and LAXPC (lower panel) light curves of the burst on 21 June, 2017 (B1). The plotted resolution is 2.3775s.}
    \label{lcb1}
\end{figure}

We detected a total of 27 bursts from 4U 1636--536 in the LAXPC data as shown in Table~\ref{laxpcbursts}. Both long and short-duration bursts were detected with different morphologies. The burst rise times and stop times correspond to the points at which the flux becomes 10 $\%$ of the maximum burst flux. We did a detailed spectral and timing study of the bursts detected in the LAXPC data. The spectral analysis includes a detailed analysis of each of the bursts using time-resolved spectroscopy and those analyses will be discussed in detail in our future work. For this work, we concentrated on a subset of the LAXPC bursts which were simultaneously covered by SXT to carry out broadband burst spectroscopy. We studied the SXT light curves corresponding to the LAXPC observations and a total of 15 bursts as shown in Table~\ref{burstsample} were detected simultaneously with LAXPC. Figure~\ref{lcb1}  (and Figure~\textbf{A1}) shows the simultaneous SXT and LAXPC light curves corresponding to the detected burst in our sample.  The burst profiles were consistent with a typical thermonuclear bursts  of different durations and morphologies \citep{2008ApJS..179..360G}. 
\subsection{Pre-Burst Spectra}

 A spectrum of 100 s intervals were extracted as the persistent spectra from the light curves starting from $\sim$ 200 s before the burst starts for each of the observations. In each case, we fitted the joint LAXPC and SXT pre-burst spectra with a model consisting of absorbed blackbody and a power-law {\tt const*(Tbabs*(bbodyrad + powerlaw)} in the 4.0--25.0 keV (LAXPC) and 0.7--7.0 keV (SXT) energy range. {\tt TBABS} model was used for interstellar extinction and the {\tt constant} takes care of the uncertainties in cross-calibrations of the two instruments. The systematic error considered was 2\%. The spectral fitting and statistical analysis were done using XSPEC v 12.10.0c spectral fitting package distributed as a part of the Heasoft 6.24 package. The best-fitting parameter values obtained were consistent but slightly lower than the previously reported literature values due to the variation in the spectral states \citep{2021MNRAS.501..168L}. 

\subsection{Time-resolved burst spectroscopy}

\begin{figure}
    \centering
     \includegraphics[width=0.45\textwidth]{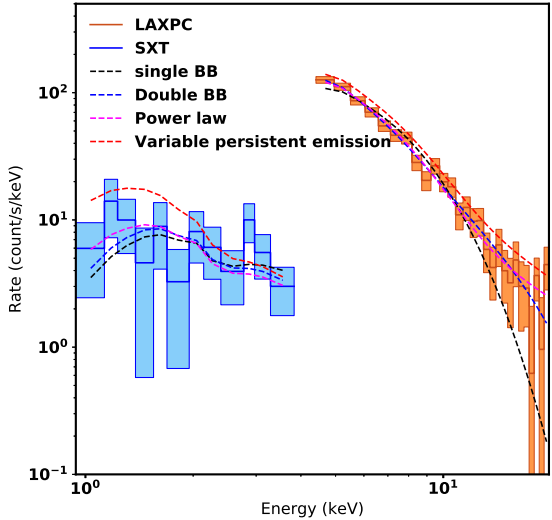}
    \caption{Comparison of different models for fitting the broadband (SXT+LAXPC) burst spectrum of a 4.7549 s segment bin (where F-test condition is satisfied) of the burst detected on 6 August 2018 (B6).  A single blackbody (black) gives a poor fit; whereas double black bodies (blue), Power-law (magenta), and a variable persistent flux (red) all give better goodness of fits to a similar extent. The shaded band represents the $1\sigma$ errors for the spectral fitting in the 0.7-20 keV band. \label{spec}}
\end{figure}

\begin{figure*}
    \centering
     \includegraphics[width=0.45\textwidth]{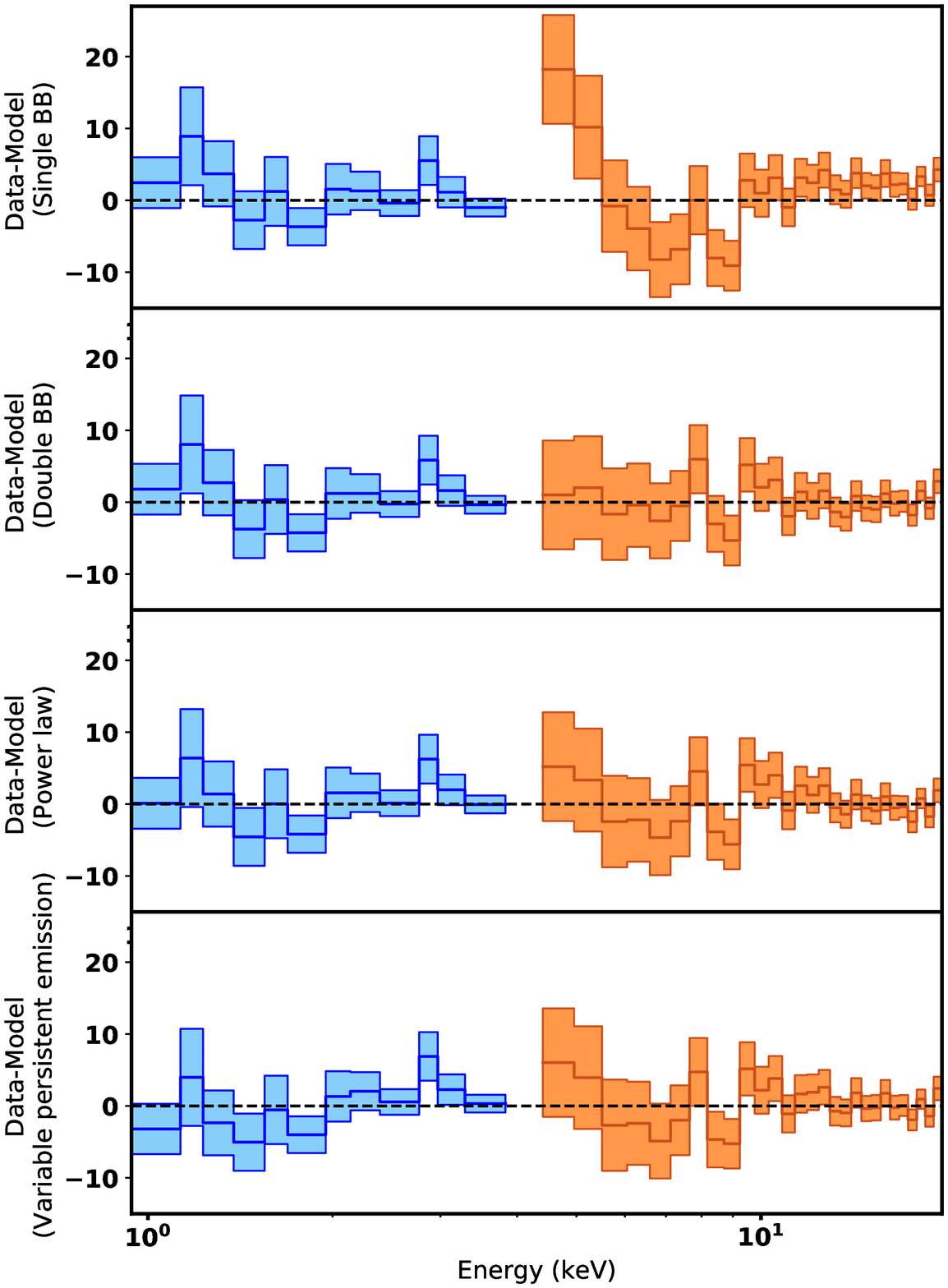}
     \includegraphics[width=0.45\textwidth]{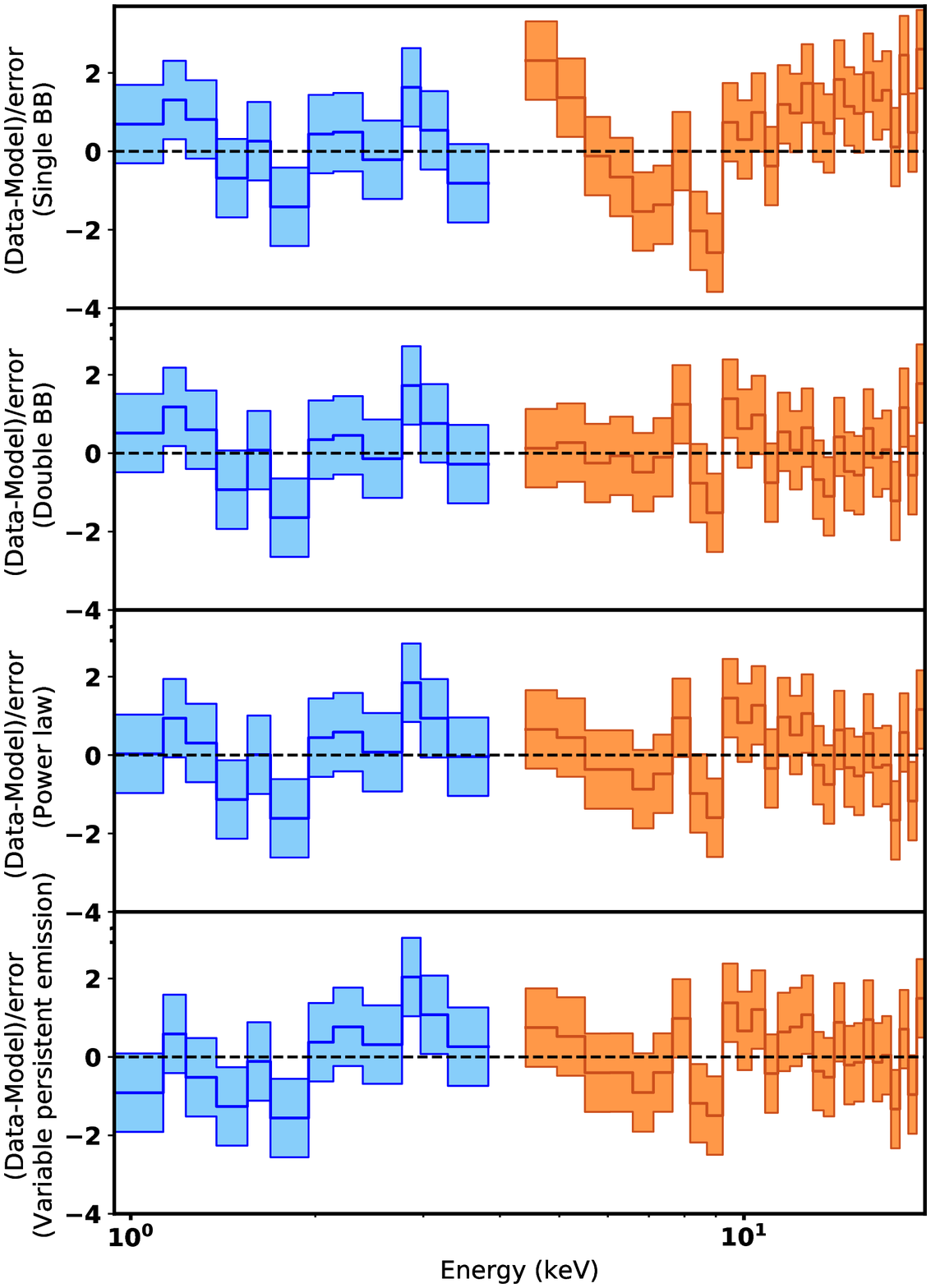}
     \caption{Residuals (left) and error-weighted residuals (right) for joint spectral fits to the 4.7549 s spectrum bin (where F-test condition is satisfied) of burst B6 from \src. A single blackbody shows strong residuals (1st panel); two blackbodies (2nd panel), blackbody+power law (3rd panel), and blackbody+a variable persistent flux (4th panel) all give better fits to similar extents. The shaded region shows the $1 \sigma$ error. The dashed line shows the zero deviation level. \label{specres}}
\end{figure*}

To probe the spectral evolution during the thermonuclear bursts, the spectra were extracted from each of 4.7549 s time interval segments during the bursts. This is because we are limited by the time-resolution of SXT which is {\bf $\sim$ 2.37 s}, and to obtain a spectrum with significant statistics (especially from SXT data) we had to extract spectrum from at least a duration of 4.7549 s. It should be noted here that to conduct joint spectroscopy, spectra were extracted from the same intervals from the LAXPC data as well. Due to low counts, spectra for bins were grouped using GRPPHA to have a minimum count rate of 10 counts per bin. Although 10 counts per bin is a fairly low grouping criterion, we still went with it to have enough grouped energy bins to enable enough degrees of freedom for the fit. We considered a Churazov weighting for the spectral fitting to take care of any effects of the non-Gaussianity of the data \citep{1996ApJ...471..673C}.
We first modeled the burst spectra with a single blackbody model  (\textbf{BB})  which is the generally followed procedure for modeling burst spectra. We fitted the burst spectra taking pre-burst spectra described in \S~3.3, as the background. The burst spectra for each time bin were fitted jointly for both LAXPC and SXT instruments in XSPEC using the model {\tt const*Tbabs*(bbodyrad)}, consisting of parameters $kT$ and normalisation, $N=(R/d_{10})^{2}$,  where R is the blackbody radius in kilometers and $d_{10}$ is the source distance in units of 10 kpc. The neutral hydrogen column density $n_{H}$ was fixed as $ 0.379 \times 10^{22}$ cm$^{-2}$ \citep{2008ApJ...688.1288P}. TBABS model \citep{2000ApJ...542..914W} was used for interstellar extinction and systematic error considered for time-resolved spectroscopy was 2\%. 
However, BB fits showed a substantial excess in the fit residuals, especially for the bins near the peak of the bursts. So, we tested three alternative two-component phenomenological models to investigate the excess by adding a second blackbody (\textbf{DBB}) {\tt const*(Tbabs*(bbodyrad + bbodyrad))}, a power-law {\tt const*(Tbabs*(bbodyrad + powerlaw))}, and a variable persistent emission component to fit the burst spectra. In the last case, we fitted the burst spectra with an absorbed blackbody model taking instrumental background (obtained from the pipeline in case of LAXPC) as the background instead of the pre-burst  spectrum \citep{2013ApJ...772...94W}  i.e., the burst spectra for each time bin were fitted in XSPEC using a model {\tt const*Tbabs*(bbodyrad) +$f_{a}$*(const*(Tbabs*(bbodyrad + powerlaw)))}. Here the constant $f_a$ is a factor quantifying the change of the persistent emission during the burst. The spectral fits were performed within the 4.0-20.0 keV energy range for LAXPC and 0.7-4.0 keV for SXT. All of these models gave better fits to the burst spectra as shown in Figure~\ref{spec}. The corresponding fit residuals and error-weighted residuals are shown in Figure~\ref{specres}. A significant excess in the fit residuals was observed when the burst spectra for the bins near the peak were fitted with the BB model.
 Whereas adding a second component (a blackbody or a power law or a variable persistent component) to the model took care of the fit excess observed in the case of the BB model making the fit better (Figure~\ref{spec}). The $\chi^2$ values \& DOFs corresponding to the fits are given in Table~\ref{tabfitstat}.  An inspection of the fit residuals confirms that the DBB model typically provides the least residuals most frequently among the two-component models, particularly for the time bins around the peak of the bursts (Figure~\ref{specres}). Hence, we use the DBB model, only as a representative of all the two-component models as all the two-component models are statistically equivalent.      

\begin{table}
\centering
\caption{$\chi^2$ and corresponding DOFs obtained from fitting the broadband (SXT+LAXPC) burst spectrum of a 4.7549 s segment bin of the burst detected on 6 August 2018 (B6) using single BB, double BB, BB+power-law, and  BB+a variable persistent emission models as displayed in Figure~\ref{spec} \label{tabfitstat}}
\begin{tabular}{lcc}
\hline
Model & $\chi^2$ & DOF \\
\hline
Single BB & 66.88 & 37 \\
BB+BB & 28.10 & 35\\
BB+power-law & 30.57 & 35 \\
BB+Variable persistent emission & 32.17 & 36 \\
\hline
\end{tabular}
\end{table}

To check the statistical significance of the best fitting spectral models corresponding to each bin during the burst, we performed an F-test taking the corresponding fit statistic ($\chi^2$) values of both BB and DBB models for each of the bins during the burst. We put a threshold of $3 \sigma$ significance on the chance probability obtained through the F-test i.e., the second component will only be considered in the cases where the chance probability of the improvement of the fit is less than 0.27\%.  We have also checked the satisfaction of the threshold criterion based on the chance probability of the improvement of the fit obtained using the \cite{2002ApJ...571..545P} methodology and except for a handful of bins in the boundary of the DBB model to the single BB model satisfaction transition, no significant deviation is observed from the results obtained using the F-test. Moreover, We observed that for most of the spectra during intervals close to the burst peak where the burst intensity was high, the F-test condition was satisfied strongly requiring the existence of the second component. 
For the brighter and longer duration bursts, the spectra of a significant fraction of the time bins during the bursts required a second component. 

\begin{figure*}
        \centering
        \begin{tabular}{lr}
         \includegraphics[width=0.49\textwidth]{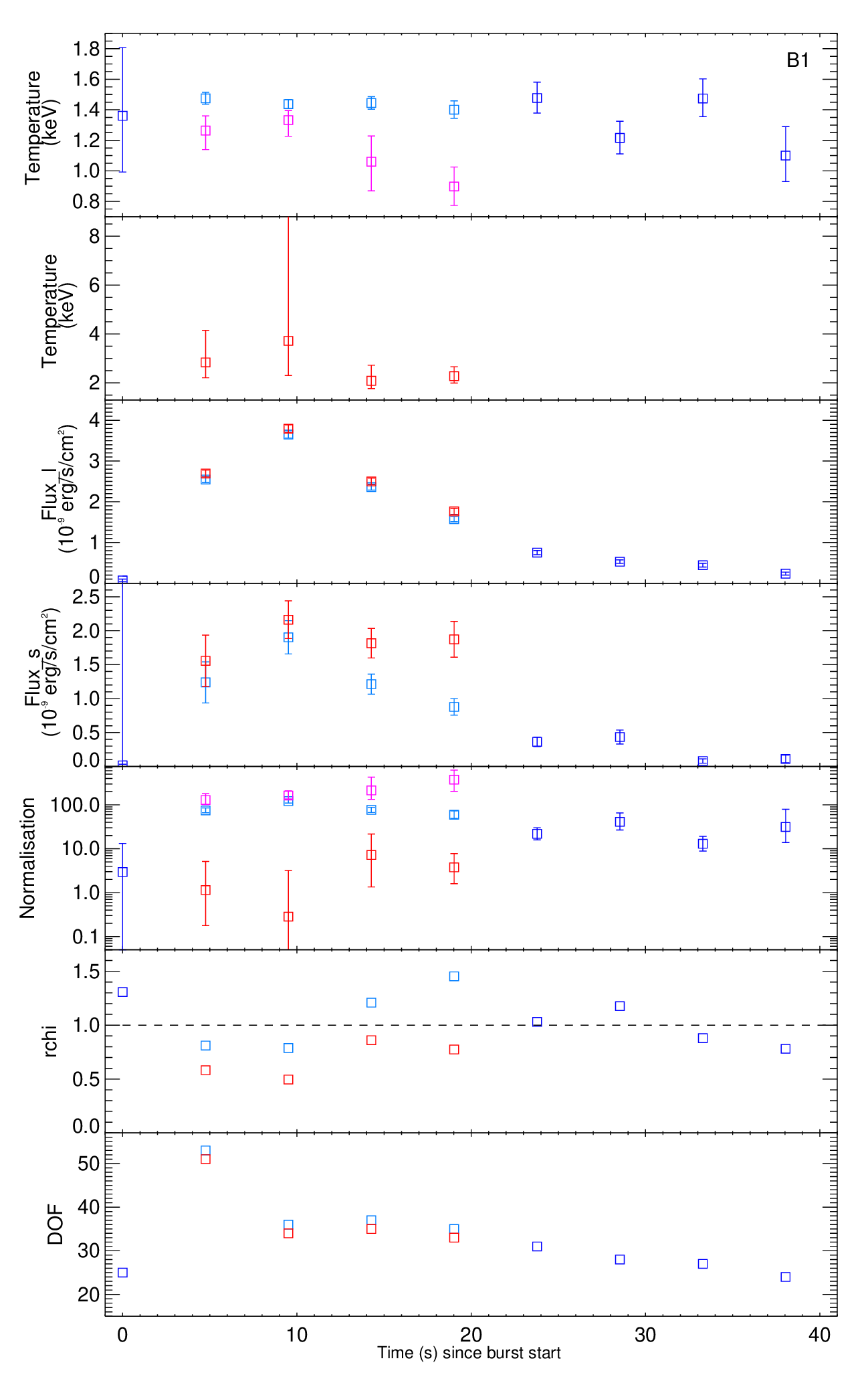} &
         \includegraphics[width=0.49\textwidth]{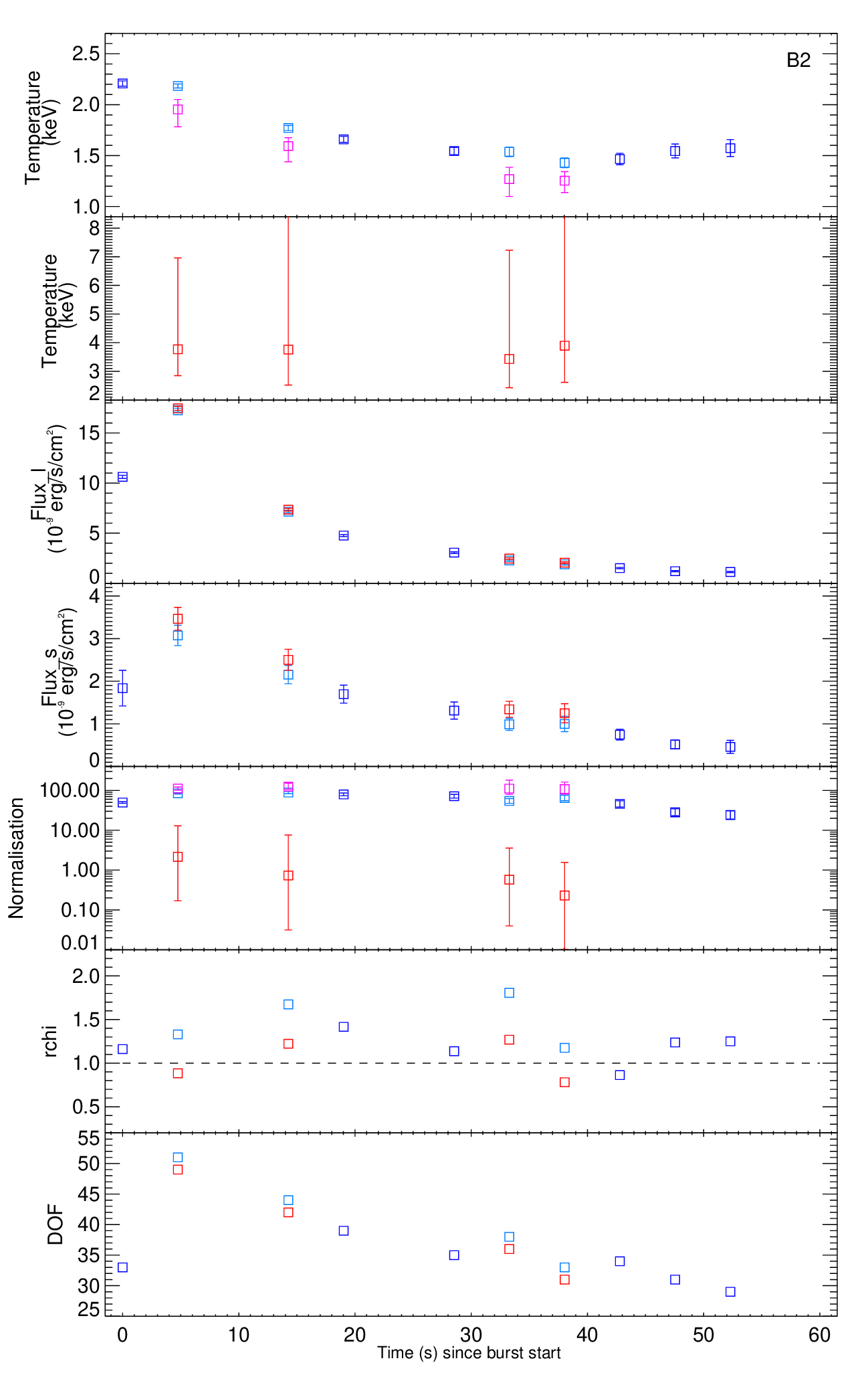} \\
         \end{tabular}
         \caption{Best fitting parameters from time-resolved spectroscopic analysis of all the 15 bursts detected with single and double blackbody models ({\bf and figure B1}). 1st panel: temperature of the single blackbody model (blue) or the temperature of the cooler component of the double blackbody model (magenta) where it is required, 2nd panel: temperature (red) of the hotter component where the double blackbody model is required, 3rd panel: 4.0--20.0 keV flux from LAXPC for single blackbody (blue) and double blackbody (red), 4th panel: 0.7--4.0 keV flux from SXT for single blackbody (blue) and double blackbody (red), 5th panel: normalisation of the single blackbody (blue) and the cooler  (magenta)/hotter (red) blackbody where double blackbody was required, 6th panel: reduced $\chi^2$ values corresponding to the fits for the single blackbody model (blue) and the double blackbody model (red) where required, 7th panel: DOFs corresponding to the fits for the single blackbody model (blue) and the double blackbody model (red) where required. The errors given on the parameters are 1$\sigma$ values. \label{timeresb1}
        } 
\end{figure*}

\begin{figure*}
    \centering
    \begin{tabular}{cc}
      \includegraphics[width=0.5\textwidth]{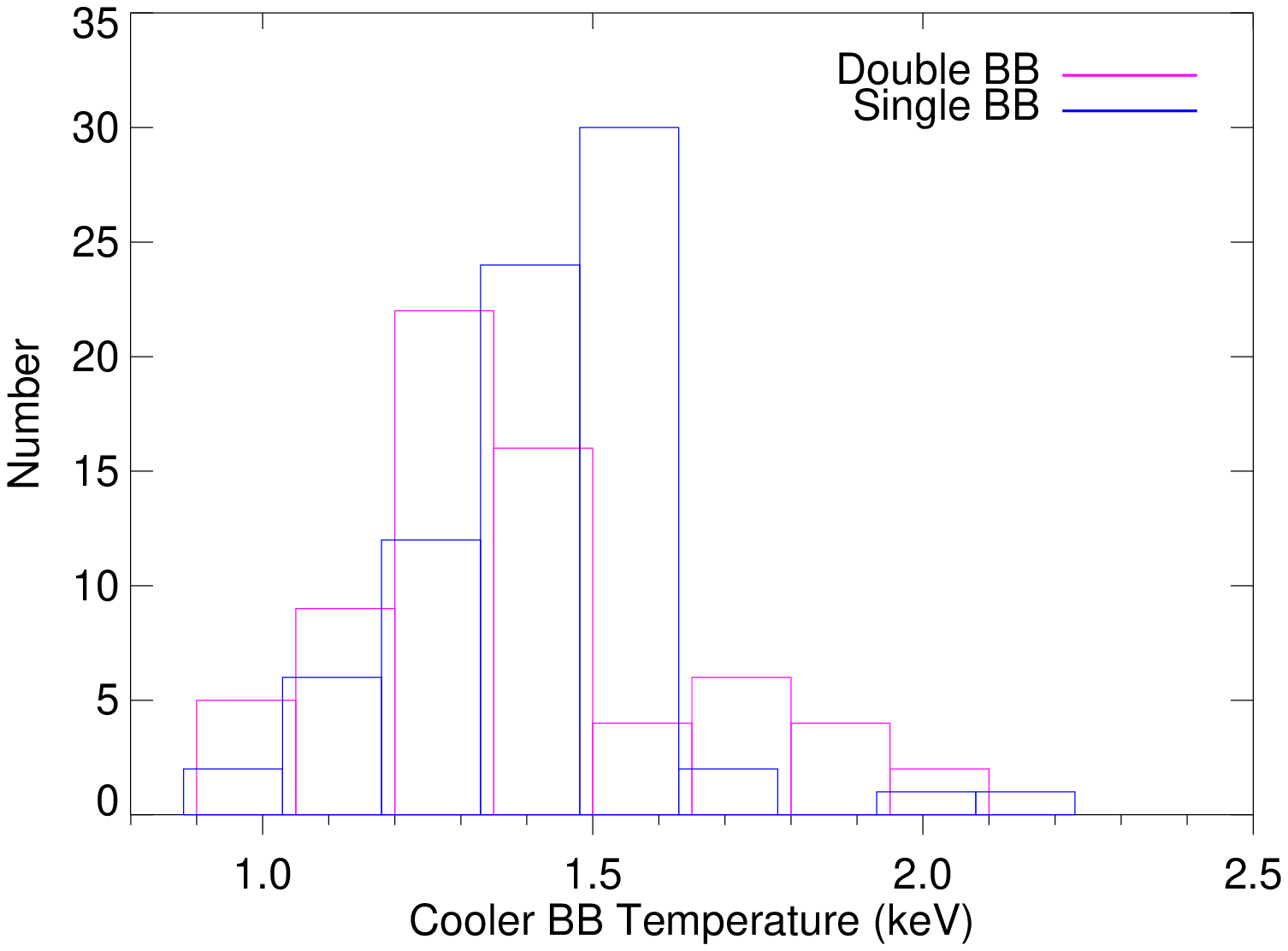} &
      \includegraphics[width=0.5\textwidth]{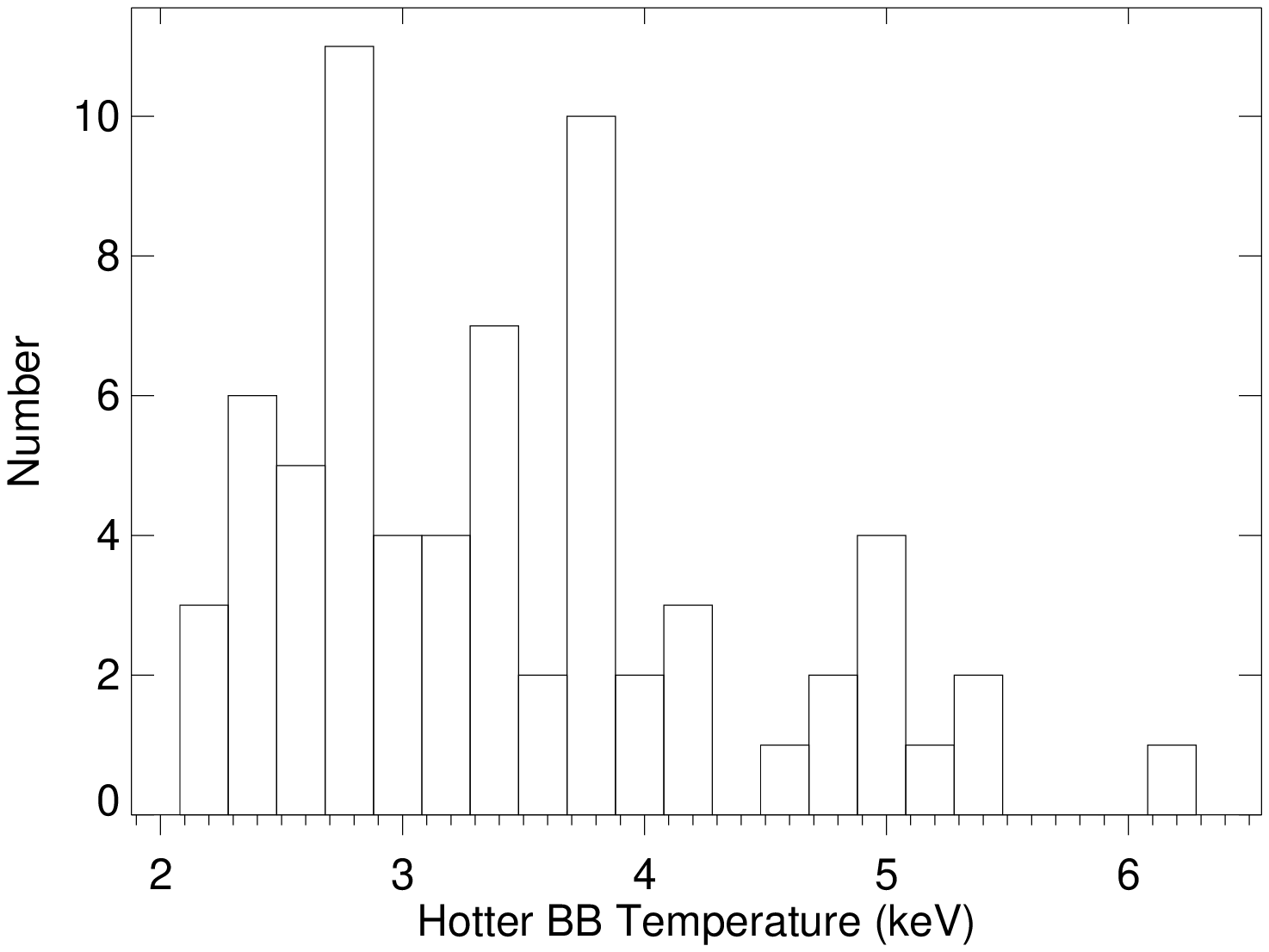}
    \end{tabular}
     \caption{Left: The number distribution of the temperature of the single BB (blue) and the temperature of the cooler component of the DBB model (magenta). Right: The number distribution of the temperature of the hotter component of the DBB model.}
     \label{temp_dist}
\end{figure*}

Figure~\ref{timeresb1}  (and Figure~\textbf{B1}) shows the time evolution of the best fit spectral parameters during the bursts in our sample.  For the time bins where the F-test condition was satisfied we show both BB and DBB model fit results. The first two panels show the evolution of temperatures during the bursts. The temperatures obtained from BB fit and the corresponding cooler temperatures obtained from DBB fits are shown in the first panel. The corresponding hotter temperatures of the DBB fits are shown in the second panel. The temperatures in the BB model and the cooler temperatures in the DBB model varied 1-2 keV whereas the hotter temperatures for the DBB model typically exhibited values above 2 keV.  The third and fourth panels show how the LAXPC (4.0-20.0 keV) and SXT (0.7-4.0 keV) flux varies during the bursts. The evolution of the normalisations obtained from the BB model and the DBB model (where it is present) are shown in the fifth panel. The evolution of both the BB normalisations and that of the cooler component of the DBB  normalisations are observed to be similar. It is to be noted that though the DBB model was very strongly required, for bins with low flux the second blackbody parameters could not always be constrained well resulting in large uncertainties in the second blackbody parameters. The reduced chi-square values corresponding to the BB and DBB (where required) are shown in the sixth panel. The seventh panel shows the DOFs corresponding to the BB and DBB (where required). It can be seen from the time-resolved spectroscopic analysis plots that the DBB model gave statistically better fits (shown in red) near the burst peaks than the BB model (shown in light blue). For the time bins nearing the end of the bursts, the BB model is sufficient to provide statistically good fits as exhibited by the corresponding $\chi^2$ values (shown in deep blue).
All the bursts were observed to show a sharp rise and exponential-like decay in flux (both LAXPC and SXT) with a cooling tail except the bursts B1, B4, B5, B8, and B10, where the cooling tails were not detected. The cooler temperatures were observed to maintain a stable trend during these bursts. Figure~\ref{temp_dist} shows the number distribution of the BB temperature, the cooler components of the DBB temperatures (left), and the hotter components of the DBB (right). The temperature distribution of the cooler component of the DBB was observed to be consistent with the BB temperature distribution whereas the distribution of the hotter component of the DBB was observed to be different - slightly skewed - exhibiting values above 2 keV.

\begin{figure}
    \centering
      \includegraphics[width=0.5\textwidth]{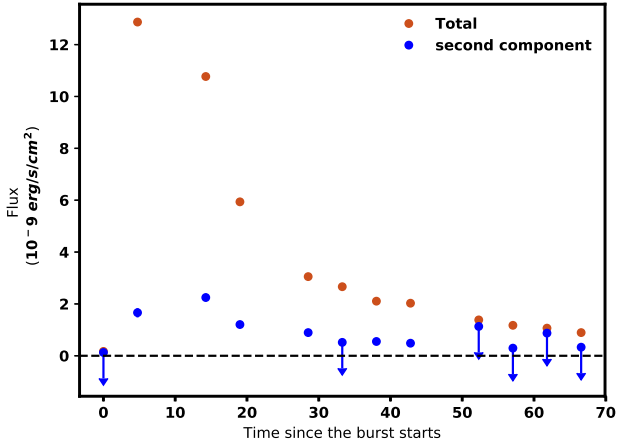}
     \caption{ The time evolution of the burst flux (4.0-20.0 keV) obtained from time-resolved spectroscopy of the burst B12.  The orange points show the total burst flux and  $1 \sigma$ errors are smaller than the graph point sizes. The blue points represent the contribution from the second blackbody in the DBB model, $1 \sigma$ errors are are smaller than the graph point sizes for the bins where F-test condition was satisfied and upper limits are shown for the rest. 
      \label{timeresfluxvar}} 
\end{figure}

Figure~\ref{timeresfluxvar} shows an example of the comparative time evolution of fluxes within 4.0-20.0 keV obtained for the burst B12 using {\tt const*(Tbabs*(bbodyrad + bbodyrad))} model. The total flux is given in orange and whereas the blue curve represents the flux contribution of the additional component of the DBB model.  As the burst decays, the absolute value of second BB intensity decreases. So, as a consequence, the observable impact of the second BB component is more pronounced near the intervals around the peak of the burst. A weaker contribution of the additional component was also observed for the relatively fainter bursts suggesting a dependence of the contribution of the second component on the strength of the total flux.

 It is to be noted here that for ensuring significant statistics we have considered relatively larger time bins (4.7549 s) for broadband time-resolved spectroscopy. To check the effect of this averaging, we generated LAXPC spectra with finer time ($\sim$ 1s) resolutions corresponding to the joint time-resolved spectroscopy bins for these bursts. These LAXPC spectra were fitted with {\tt tbabs*bbodyrad} model individually. The average of the burst parameters from these only LAXPC spectra was found to be generally consistent with the burst parameters obtained from joint spectroscopy at the same time.  Finally, we compared the combined BB model {\bf {\tt tbabs*(bbodyrad+bbodyrad+bbodyrad+bbodyrad+bbodyrad)}} (governed by the parameters obtained from the fine LAXPC time-resolved analysis) for the corresponding joint spectra with the residuals obtained from our joint spectral fitting. The presence of significant fit residuals for the former compared to the latter shows that our findings are not primarily due to the contribution of the time averaging of the spectra and is far beyond what is expected from the effect of averaging over larger bins.

\clearpage
\clearpage

\clearpage

\section{Discussion}

In this paper, we report the detection of 15 thermonuclear bursts covered simultaneously with LAXPC and SXT onboard AstroSat. We analyzed the broadband data of all the 15 Type I X-ray bursts. We report and discuss the results obtained from time-resolved spectroscopy of the bursts and investigate the physical interpretations of the excess observed near the peak of the bursts. We also studied the dependence of the hotter BB temperature and its requirement for a good spectral fit on the varying state of the source as observed from HID (Figure~\ref{HID}).  But no such state dependence was observed although the HID showed a modest spectral evolution within the island state of the atoll track \citep{2008ApJ...685..436A}. Also, the properties of the bursts including duration, morphology, burst peak flux as well as burst fluence were observed to exhibit no systematic trend and displayed significant variation even within the same observation. Moreover, we studied the energy-dependent burst light curves of all the 15 bursts reported and  no hard X-ray dips implying coronal changes could be detected in our burst sample as the data presented here does not have sufficient sensitivity to detect hard X-ray dips of the magnitude reported by \cite{2014ApJ...782...40J}.

From the broadband time-resolved spectroscopy of all the simultaneously detected Type I bursts in our sample, it was observed that the time intervals near the burst peaks were well described by two-component models as compared to the BB models usually used for describing burst spectra. We have demonstrated that with broadband coverage the deviation from the widely used Planckian spectrum of the burst \citep{2020MNRAS.499..793B, 2018ApJ...855L...4K, 2018ApJ...856L..37K, 2019ApJ...885L...1B} can be clearly ascertained and the burst-accretion interaction can be traced out. From our time-resolved spectroscopic results using the best fit models, the characteristic thermal cooling tail during burst decay was observed for all the bursts except B1, B4, B5, B8, and B10. For these 5 bursts, the cooler BB temperatures were observed to maintain a stable temperature value throughout the bursts. The subset of bursts with the absence of a clear cooling trend systematically corresponded to a relatively shorter duration.
The absence of cooling during X-rays does not necessarily dispute their thermonuclear origin, as such "non-cooling" (where thermal cooling is not detected) bursts have previously been detected from several bursting sources \citep{2011ApJ...733L..17L, 2021arXiv210111637G}.

\begin{figure}
    \centering
     \includegraphics[width=0.4\textwidth]{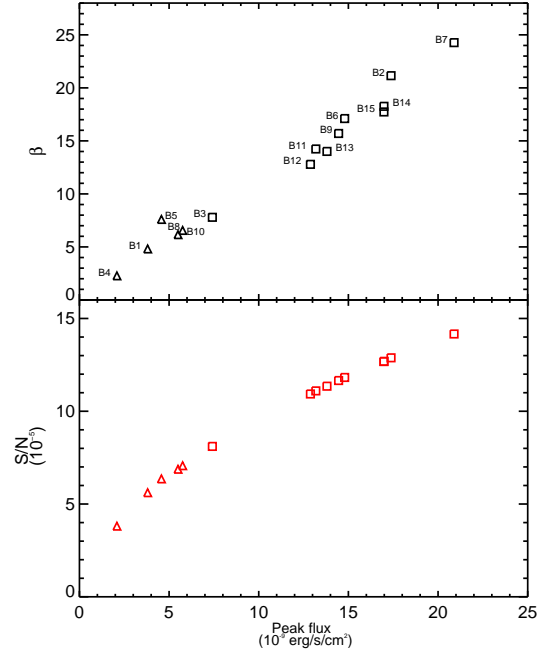}
     \caption{Evolution of ratio $\beta$ (top) and $S/N$ (bottom) with the burst peak flux in 4.0-20.0 keV. '$\triangle$' stands for the "non-cooling" bursts. }
      \label{betaevol}
\end{figure}

 To further investigate the absence of cooling in these bursts, we followed the same approach as \cite{2011ApJ...733L..17L} to examine the origin of "non-cooling" bursts. We obtained the peak-to-persistent flux ratio $\beta$ (=$F_{peak}/F_{pers}$) as well as signal-to-noise ratio ($S/N=F_{peak}/\sqrt{F_{peak}+F_{pers}}$) for all the bursts in our sample. These are figure-of-merit parameters for quantifying the strength of the burst emission over persistent emission for each burst. The values of both $\beta$ as well as $S/N$ were observed to be systematically lower for these "non-cooling" bursts in our sample (Figure~\ref{betaevol}).  Also, although a sharp transition is observed from a cooling behavior to a non-cooling behavior as a function of peak flux between B3 and B10, the observed difference between the bursts B3 and B10, based on burst fluence and duration is not very significant (Table 2). As the burst intensity decreases near the tail of the bursts for these low $S/N$ bursts, the variation of temperature and normalisation becomes less constrained and more susceptible to initial values due to a decrease in sensitivity to spectral cooling. In general, the bursts observed at different accretion rates during an outburst are observed to become less energetic with the increase in the persistent flux of the bursts due to unstable burning conditions being more quickly reached at a high accretion rate \citep{2011MNRAS.418..490C,2011ApJ...733L..17L}. The higher accretion rate may lead to a hotter neutron star photosphere which in turn may smear out the burst cooling tails \citep{2011ApJ...733L..17L}. But the accretion rate does not change much for our burst sample, and we see from Figure~\ref{betaevol} that the lower peak flux bursts show lower $\beta$ and $S/N$. Hence the "non-cooling" tails observed is due to the decrease in sensitivity to spectral cooling.

We investigate the interpretations of the excess observed in the residuals of the spectra near the peak of the bursts for the single blackbody fits. We tried to fit the spectra with three two-component phenomenological models involving thermal, non-thermal, and variable persistent emission (Figure~\ref{spec}). Similar kinds of excess were found in such LMXB sources with multiple other instruments such as NICER \citep{2020MNRAS.499..793B, 2018ApJ...855L...4K, 2018ApJ...856L..37K, 2019ApJ...885L...1B}, especially those having soft X-ray coverage. The interpretations of this excess observed from 4U 1636-536 has also been reported and discussed \citep{2013ApJ...772...94W, 2014ApJ...797L..23K, 2014HEAD...1412227K, 2016ApJ...829...91K}. Near the peak, where the burst flux was higher, all the two-component models performed significantly better suggesting a stronger deviation from the canonical Planckian burst spectrum model at these fluxes. 
The reason behind the excess may possibly be the burst emission incident on the disk and re-emitted by the disk resulting in an extra black body \citep{2018ApJ...855L...4K, 2004MNRAS.351...57B, 2004ApJ...602L.105B,2014ApJ...797L..23K}. The detection of features such as emission lines and absorption edge during the time-resolved spectral analysis of the RXTE PCA  spectrum of the 2001 superburst from  4U 1636--536 suggests reflection of the superburst radiation off the accretion disk \citep{2014ApJ...797L..23K}. Our data for the X-ray bursts were not sensitive enough to differentiate among the possible interpretations of the two-component models as well as for possible detection of such features using such complex models. 
However, photons may also be scattered in the neutron star atmosphere. The hot neutron star atmosphere may scatter the burst photons coming from the hot-spot and increase the observed temperature and spectrum \citep{1987ApJ...313..718R,1986ApJ...306..170L,2004ApJ...602..904M,2011A&A...527A.139S,2012A&A...545A.120S}. This is usually combined in the color factor use to infer the intrinsic temperature and radius from the effective observed temperature and radius. Furthermore, the excess may be produced by reprocessing the photons in an optically thin medium such as corona   \citep{2018ApJ...855L...4K}.  The soft excess can also be explained by an enhanced persistent component i.e., an increased persistent emission signifying an enhanced accretion rate because of the radiation torques on the accretion disk around the star by the burst photons (Poynting–Robertson drag) \citep{2013ApJ...772...94W, 2015ApJ...801...60W}. In our work we compared different simple models for the excess however, given the quality of the data, no model could be statistically favoured over the others.  

 During the strong PRE (Photospheric radius Expansion X-ray burst) burst  detected from 4U 1820-30 near its Eddington limit, a strong distinctly higher rise in the $f_a$ value was observed \citep{2018ApJ...856L..37K}. Physically, this corresponded to an additional Comptonization component during the PRE phase followed by re-brightening of the thermal component with a drop in the Comptonization flux. During the PRE burst, the NS photosphere expands to a significant radius ensuring a much stronger likelihood of interaction of burst radiation with the surrounding environment and thus may help to distinguish the fraction of the spectral excess contributed by the thermal and the non-thermal components and understanding their origin. However, the evolution of the burst temperature and normalisation as observed from the time-resolved spectroscopy suggests that none of the bursts in our sample is a PRE burst. In addition to that, the maximum peak burst flux found in our sample was 2.09 $\times 10^{-8}$ $erg/s/cm^{2}$  (4.0--20.0 keV) whereas the reported average PRE burst peak flux for 4U 1636--536 is 5.37 $\times 10^{-8}$ $erg/s/cm^{2}$ \footnote{PIMMS predicted flux of 6.4 $\times 10^{-8}$ $erg/s/cm^{2}$ RXTE flux (2.5-20 keV)} (4.0--20.0 keV)  \citep{2006ApJ...639.1033G}. This flux comparison furthermore supports the non-PRE nature of the bursts in our sample. 

\begin{figure}
    \centering
     \includegraphics[width=0.47\textwidth]{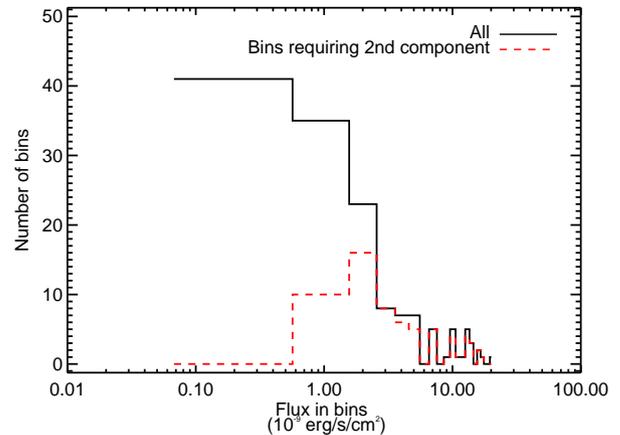}
     \caption{Distribution of observed flux in 4.0-20.0 keV in time bins.
     The black line shows the distribution of the total flux whereas the red dashed line shows the distribution of flux in time bins where the DBB model was statistically required.}
     \label{histflux}
\end{figure}

At this current juncture with the current data sensitivity, it is difficult to favor any of these models over the others and consequently any of the physical scenarios over the others. However, for the burst spectra, the double blackbody model performed the best though this can not be statistically substantiated over the other two-component models. We have also studied how the requirement of the second BB component for a good spectral fit depends on the total observed flux during the bursts, burst duration, and burst peak flux. 
Figure~\ref{histflux} shows the comparison of the distribution of observed total flux in all burst time bins along with the flux distribution (in red) in time bins where the requirement of the DBB component was statistically significant. It can be clearly observed that at low fluxes the single BB components suffices whereas as the flux increases the DBB is required in most of the burst time bins. To further quantify the significance of the dissimilarity of these two distributions we have carried out a Kolmogorov–Smirnov (KS) test. The KS test shows that the probability that two sets of flux (total flux and flux of the second BB component) distributions belong to the same parent distribution is $9.61\times10^{-8}$.

\begin{figure*}
    \centering
    \begin{tabular}{cc}
     \includegraphics[width=0.47\textwidth]{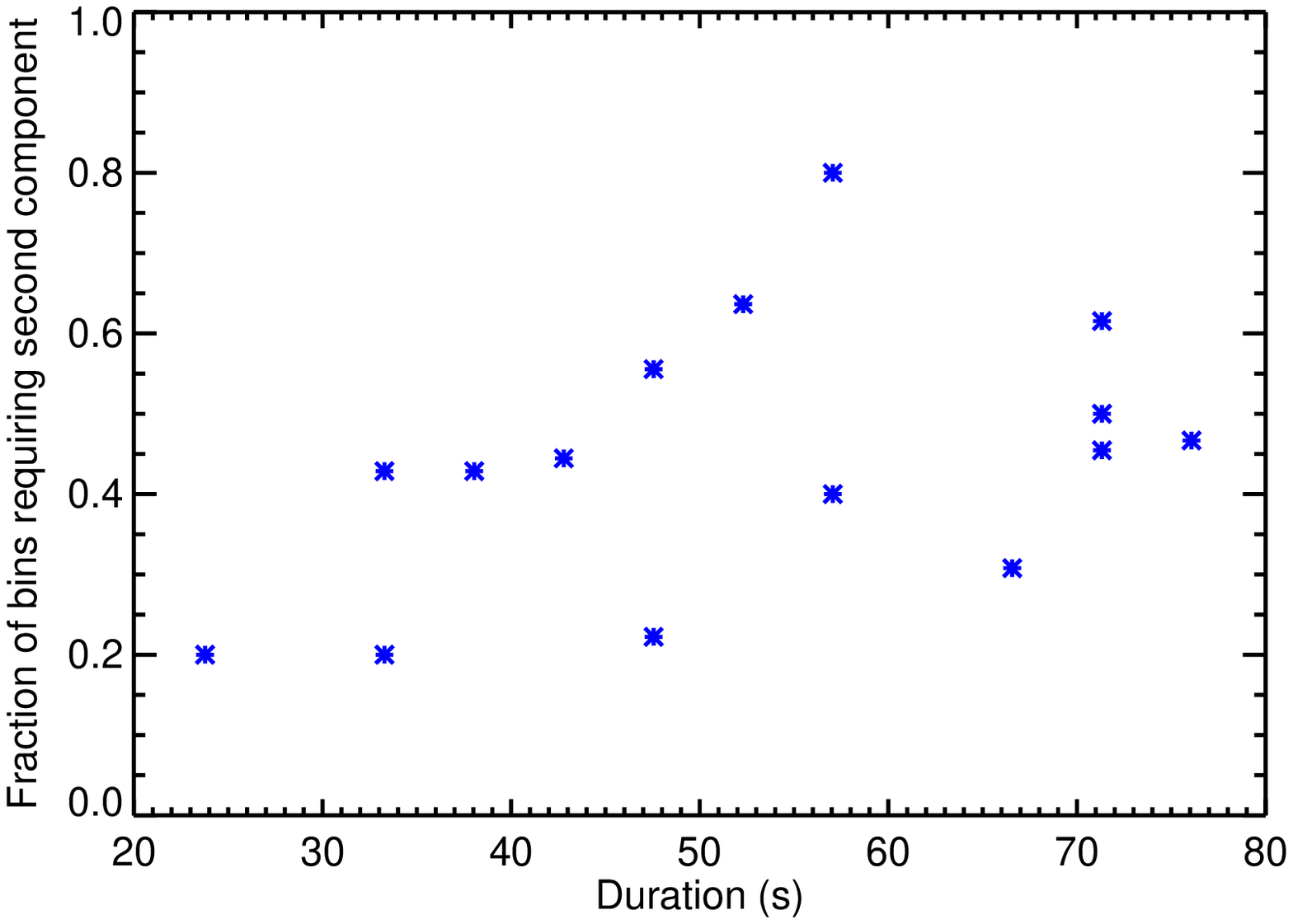} &
    \includegraphics[width=0.47\textwidth]{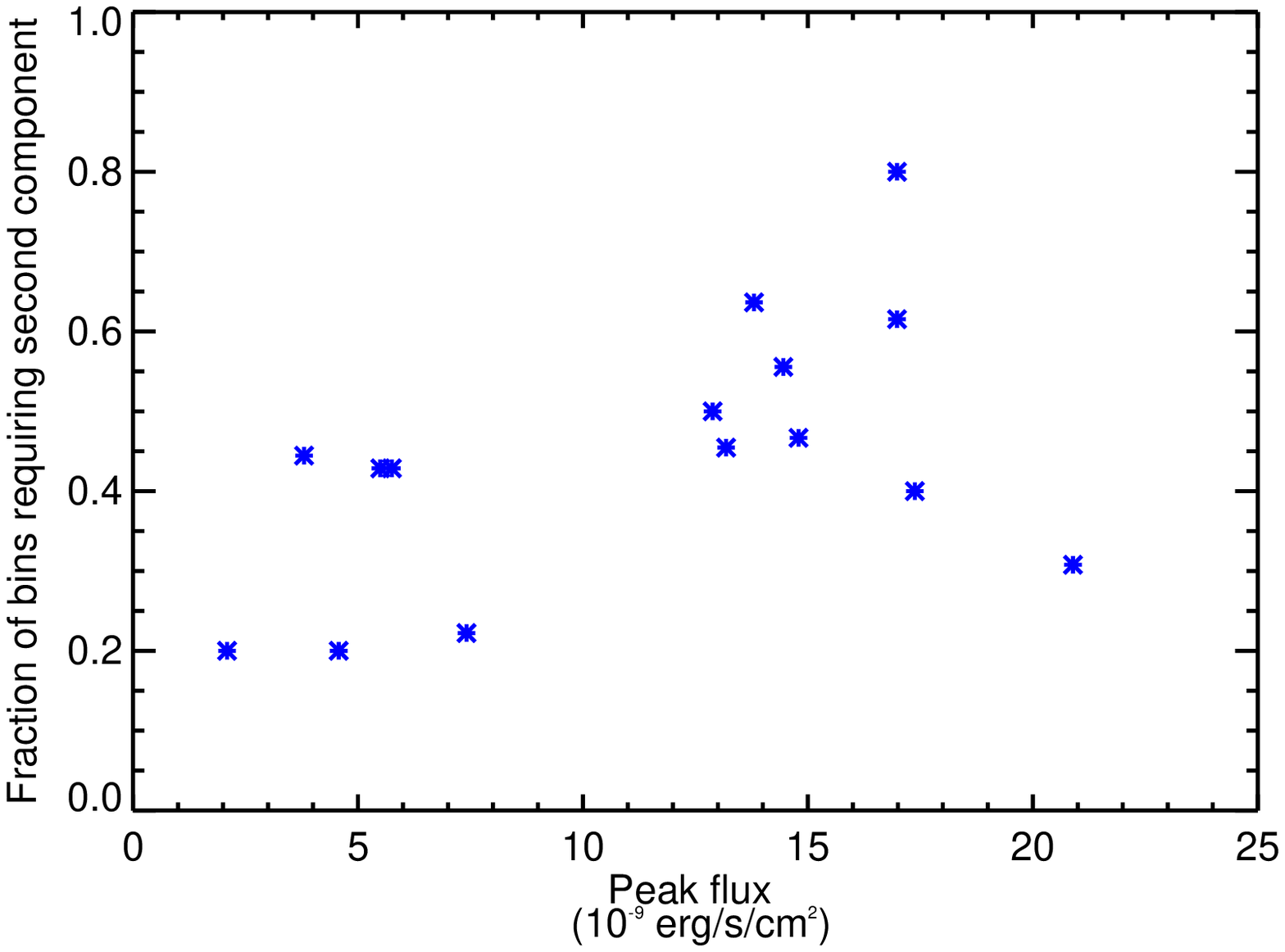}
    \end{tabular}
     \caption{Variation of the fraction of time bins requiring the second BB (from the DBB model) with burst duration (left) and burst peak flux in 4.0-20.0 keV range (right).
     }
    \label{frac}
\end{figure*}
Furthermore, the fraction of burst time bins with excess observed in the fit residuals were more for long-duration bursts as well as for the bursts with higher peak flux as shown by the left and right plots, respectively, of Figure~\ref{frac}. But, the burst B7 represents an outlier in the correlation between the peak flux and fraction of bins requiring the DBB model. This can be explained by the relatively faster burst intensity decay observed in the case of this particular burst (Figure~\ref{B7}), which occurred when the source was in a relatively harder state (Figure~\ref{HID}). A similar kind of fast decay in burst flux was observed from 4U 1636--536 earlier in its H-rich bursts in hard state revealing the impact of accretion on the bursting properties \citep{2017A&A...604A..77K}. A correlation test performed between the fraction of the total number bins requiring a second BB component for a good fit and the burst duration as well as burst peak flux show a Spearman’s rank correlation coefficient value of 0.54 and 0.73 respectively, corresponding to more than 2$\sigma$ and 3$\sigma$ significance.  It indicates that the absolute contribution of the second BB component becomes higher for the brighter and the long-duration bursts, especially near the peaks. It implies a stronger burst accretion interaction, which may also be arising from the increasing S/N ratio near the peaks of the bursts.

\begin{figure}
    \centering
     \includegraphics[width=0.47\textwidth]{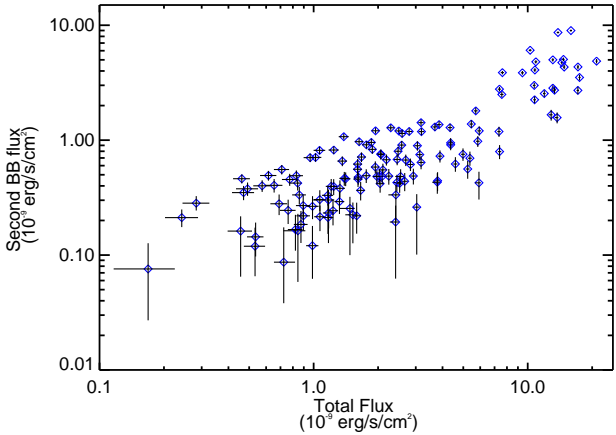}
     \caption{Variation of the flux in 4.0-20.0 keV range contributed by the additional component of DBB model with the total flux in 4.0-20.0 keV from the DBB model for all time bins during the bursts in the sample.}
     \label{corrflux}
\end{figure}

Additionally, we examined the dependence of the contribution of the second additional blackbody on the total flux at that time during the burst. Figure~\ref{corrflux} shows that the flux of the second black body is correlated with the total flux of the bursts. We found a correlation between both the fluxes with a Spearman’s rank correlation coefficient of $0.80$ corresponding to a $11 \sigma$ significance. Moreover, a  Pearson correlation coefficient computed between the two fluxes was found to be 0.83.  This dependence is observed to show a transition at $\sim$ $8 \times 10^{-9}$ $erg/s/cm^{2}$. Above this transition, the fractional contribution of the second blackbody appears to grow as the total flux increases. Whereas, below this transition, the second BB is observed to contribute about 1/6th of the total flux Figure~\ref{corrflux}. This suggests that the presence and strength of the additional component, possibly suggesting a burst-surrounding interaction, is strongly guided by the burst flux. This is consistent with the above theoretical scenarios in each of which a stronger burst flux implies a stronger interaction.  However, we did not observe any strong dependence of the hotter blackbody temperature on the total burst flux though it should be noted that given the data the second BB parameters have relatively larger uncertainties. Furthermore, the original sites of the second blackbody component may be quite complex, and a simplistic single blackbody may not be able to completely capture the nuances of the undergoing physical process.

\begin{figure}
    \centering
     \includegraphics[width=0.50\textwidth]{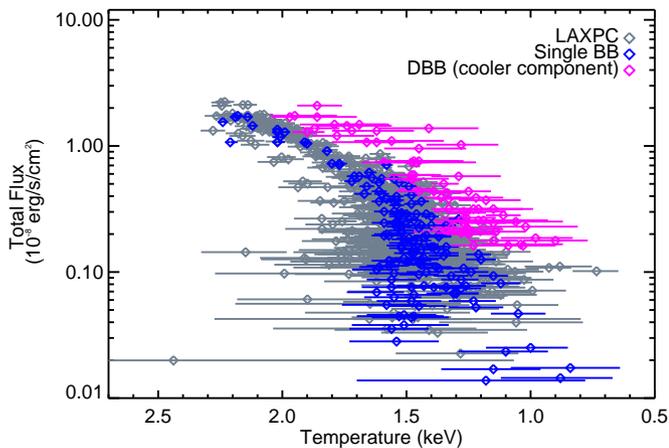}
     \caption{Variation of the blackbody temperature obtained from BB model fit of just the LAXPC spectra (grey), the single BB (blue) temperature obtained from joint (LAXPC+SXT) spectral analysis, and the temperature of the cooler component of the DBB model (magenta) obtained from the joint (LAXPC+SXT) spectral analysis with the corresponding 4.0-20.0 keV fluxes. Note that the blue and the grey overlap whereas the magenta population shows a systematic shift towards the lower temperature.}
     \label{tempflux}
\end{figure}

Figure~\ref{tempflux} shows the variation of the burst blackbody temperature with the corresponding burst flux. Here the burst temperatures from the single blackbody and double blackbody fits are shown in blue and magenta respectively. We have also carried out time-resolved spectroscopy from just LAXPC data alone with finer (1 s) time bins for these bursts, and the corresponding temperatures are shown in grey. As expected the higher-energy sensitive LAXPC fits corresponded more closely with the less effective single blackbody model temperatures, whereas with joint spectroscopy, especially at higher fluxes, DBB model was the favored and the consequent burst temperatures exhibited a systematic shift towards lower values compared to the previous two temperature populations. With NICER, which has soft energy sensitive spectral capability, a similar but opposite, relatively higher trend has sometimes been observed for the burst temperature, with burst temperatures obtained from the time-resolved spectroscopy using $f_{a}$ method showing a systematic shift towards higher values (Guver et al., in preparation). Thus our study reveals the need for broadband spectroscopic analysis of bursts. Further such studies with more sensitive broadband spectral coverage and detailed time-dependent burst emission models will be able to shed light on burst-accretion environment interaction.

\section{Conclusions}
We have presented the first broadband time-resolved spectroscopy study of 15 Type I X-ray bursts detected using LAXPC and SXT onboard AstroSat. In our broadband analysis of Type I bursts, an excess near the peaks of the bursts was observed in almost all the bursts detected simultaneously even in the relatively fainter bursts. This kind of broadband joint analysis of thermonuclear bursts provides a better platform for studying excess and deviation from the Planckian spectrum, which is not often identifiable statistically. 
It also manifests the requirements of more such broadband observations with better sensitivity.


\section*{Data Availability}
This paper includes data collected by the AstroSat mission, which is publicly available from the ISRO Science Data Archive for AstroSat Mission and the data is made available to the users by the  Indian Space Science Data Centre (ISSDC), ISRO. [https://astrobrowse.issdc.gov.in/astro\_archive/archive/Home.jsp]



\bibliographystyle{mnras}
\bibliography{ref.bib} 




\appendix
\onecolumn
\newpage 
\section{Light curves of thermonuclear X-ray bursts detected simultaneously by LAXPC and SXT onboard AstroSat}

\begin{figure*}
    \centering
     \includegraphics[width=0.40\textwidth]{Figure2.ps} 
     \includegraphics[width=0.40\textwidth]{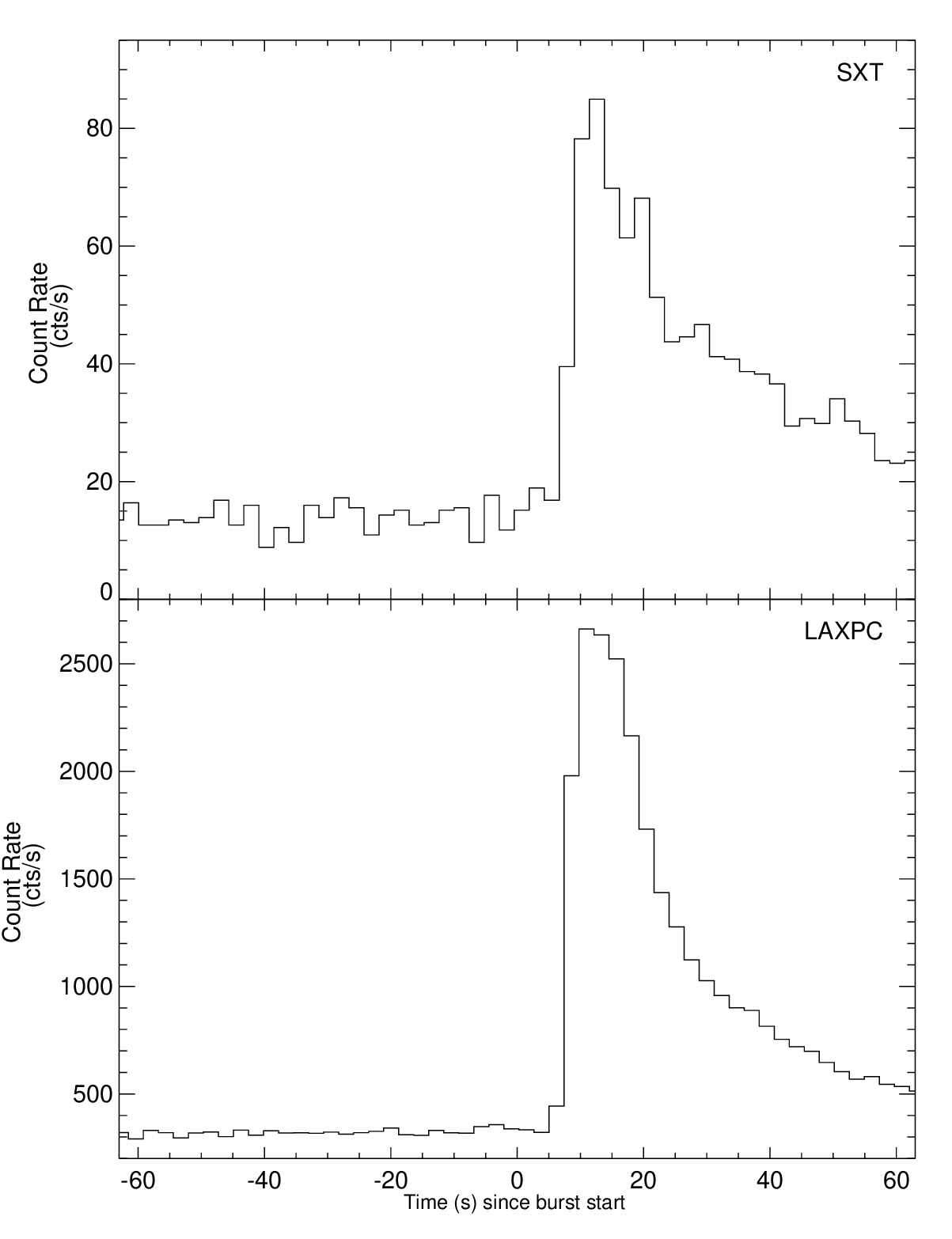} 
     \includegraphics[width=0.40\textwidth]{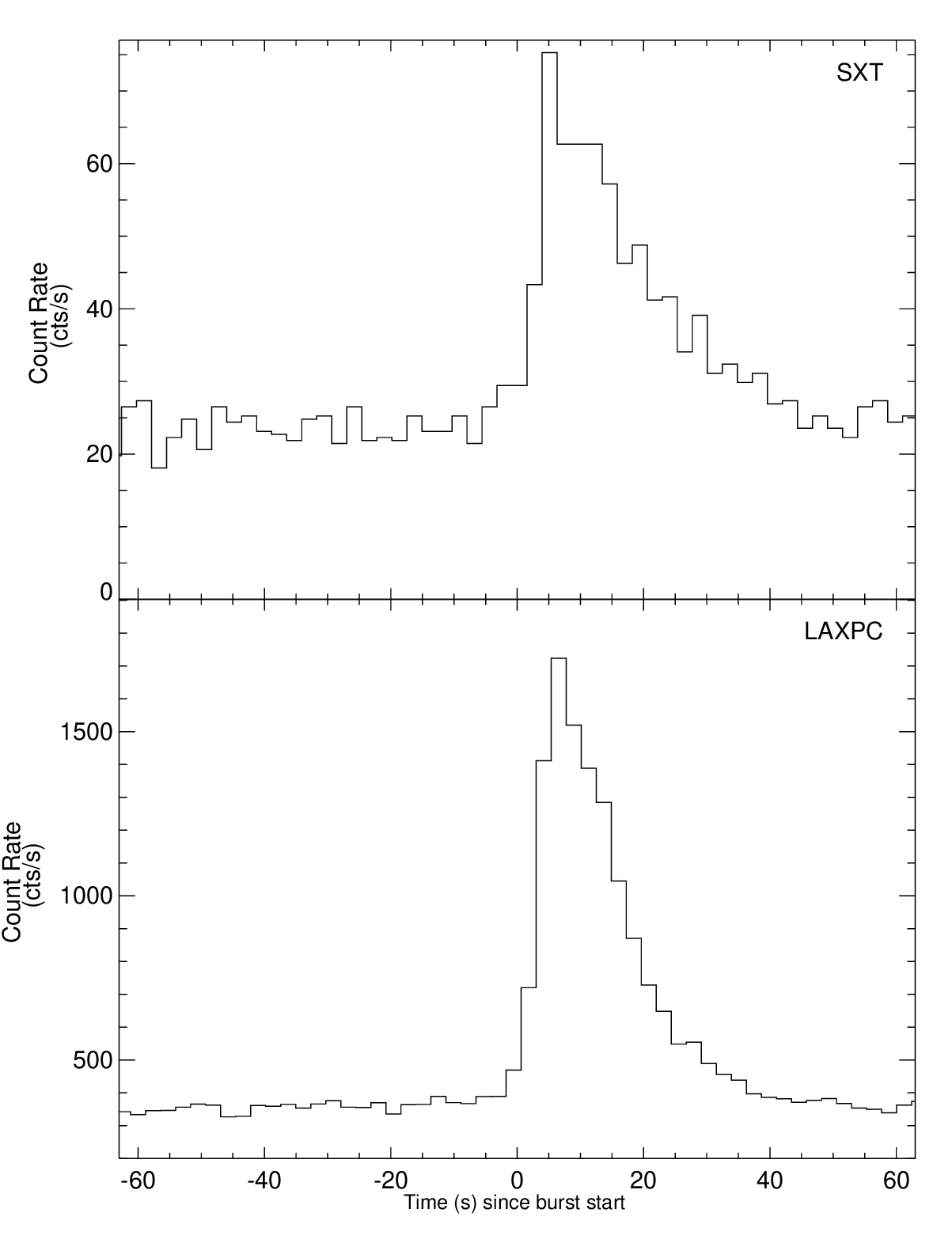} 
     \includegraphics[width=0.40\textwidth]{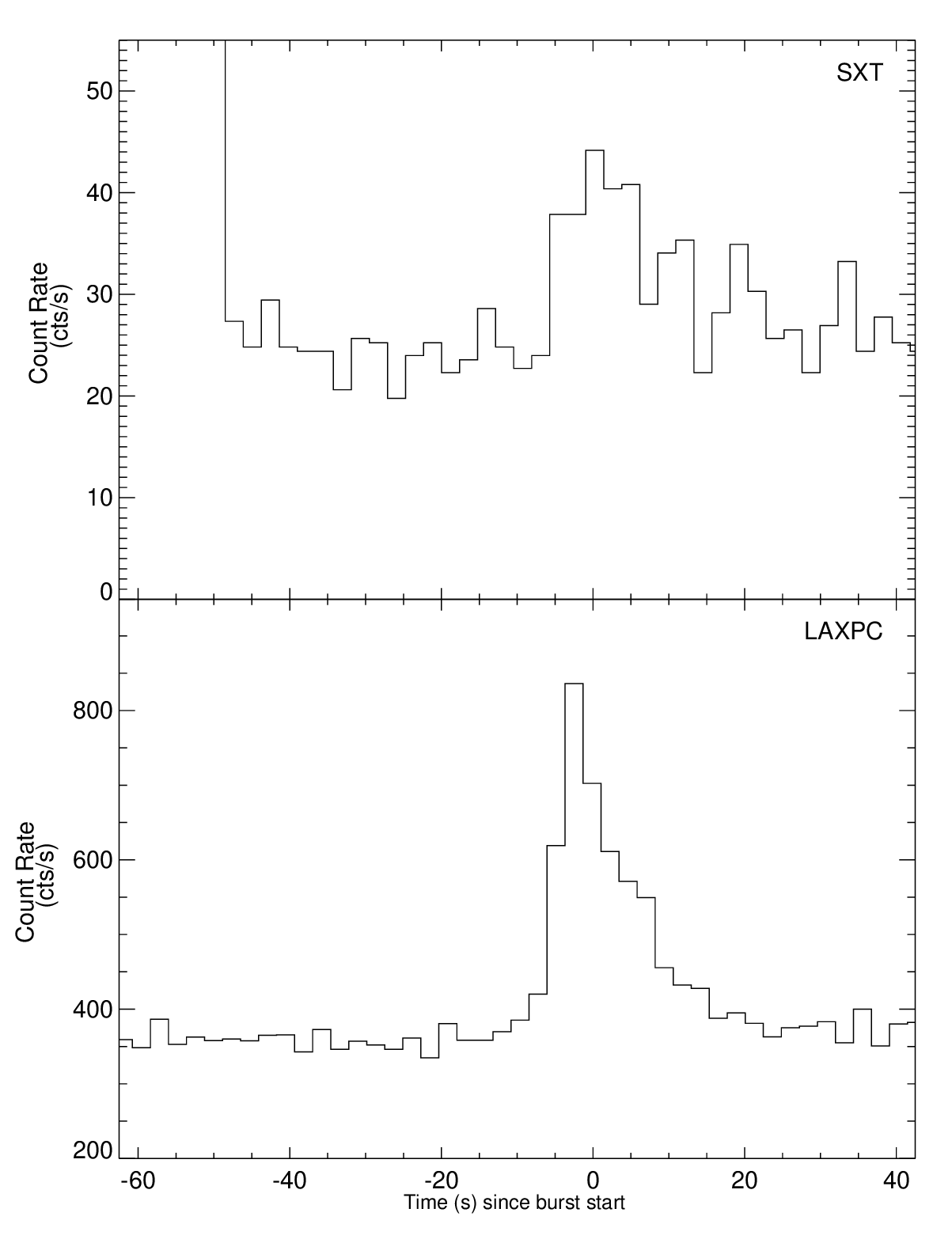} 
     \caption{Top left: B1, Top right: B2, Bottom left: B3, Bottom right: B4}
  
\end{figure*}

\addtocounter{figure}{ -1}
\begin{figure*}
    \centering
     \includegraphics[width=0.40\textwidth]{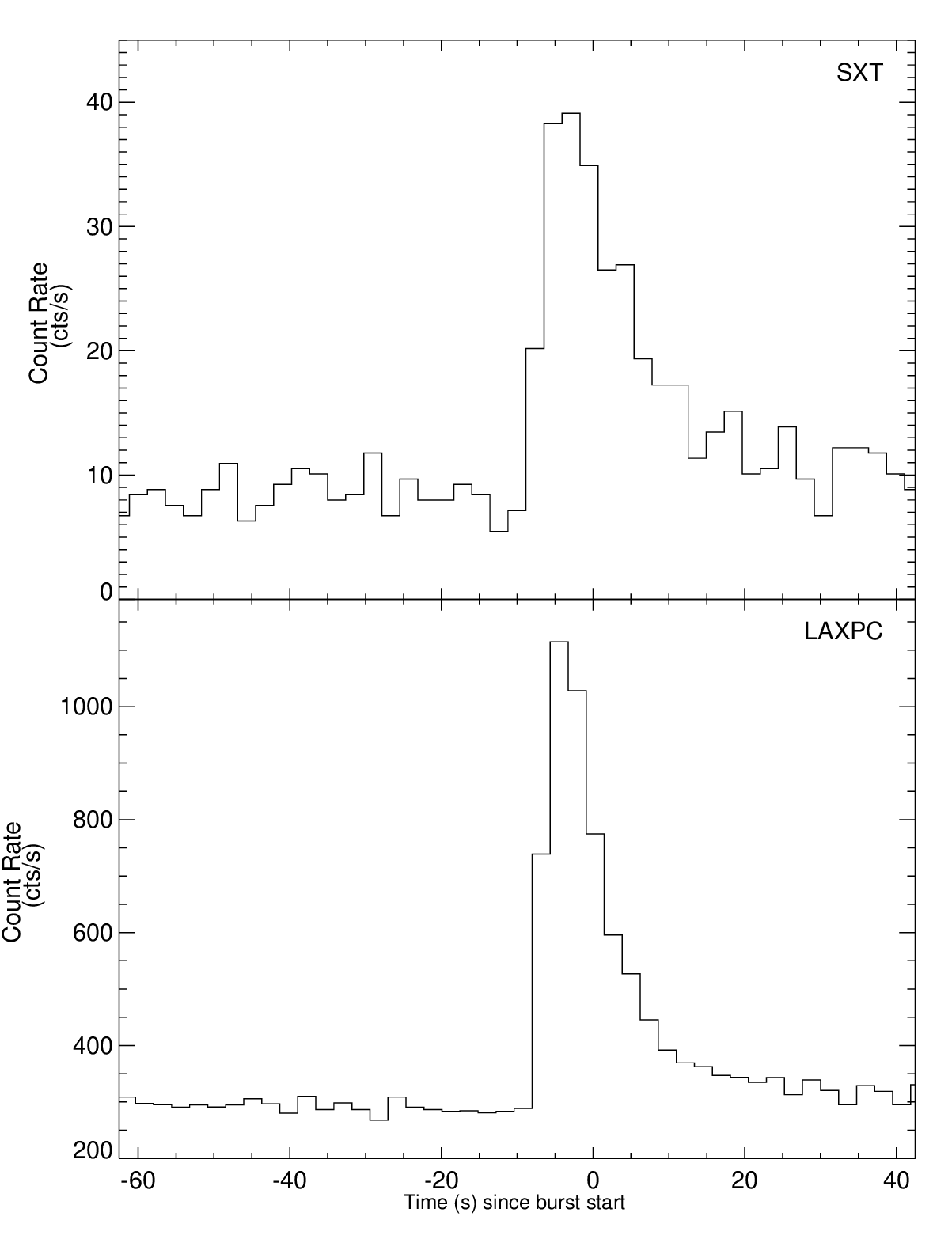}
     \includegraphics[width=0.40\textwidth]{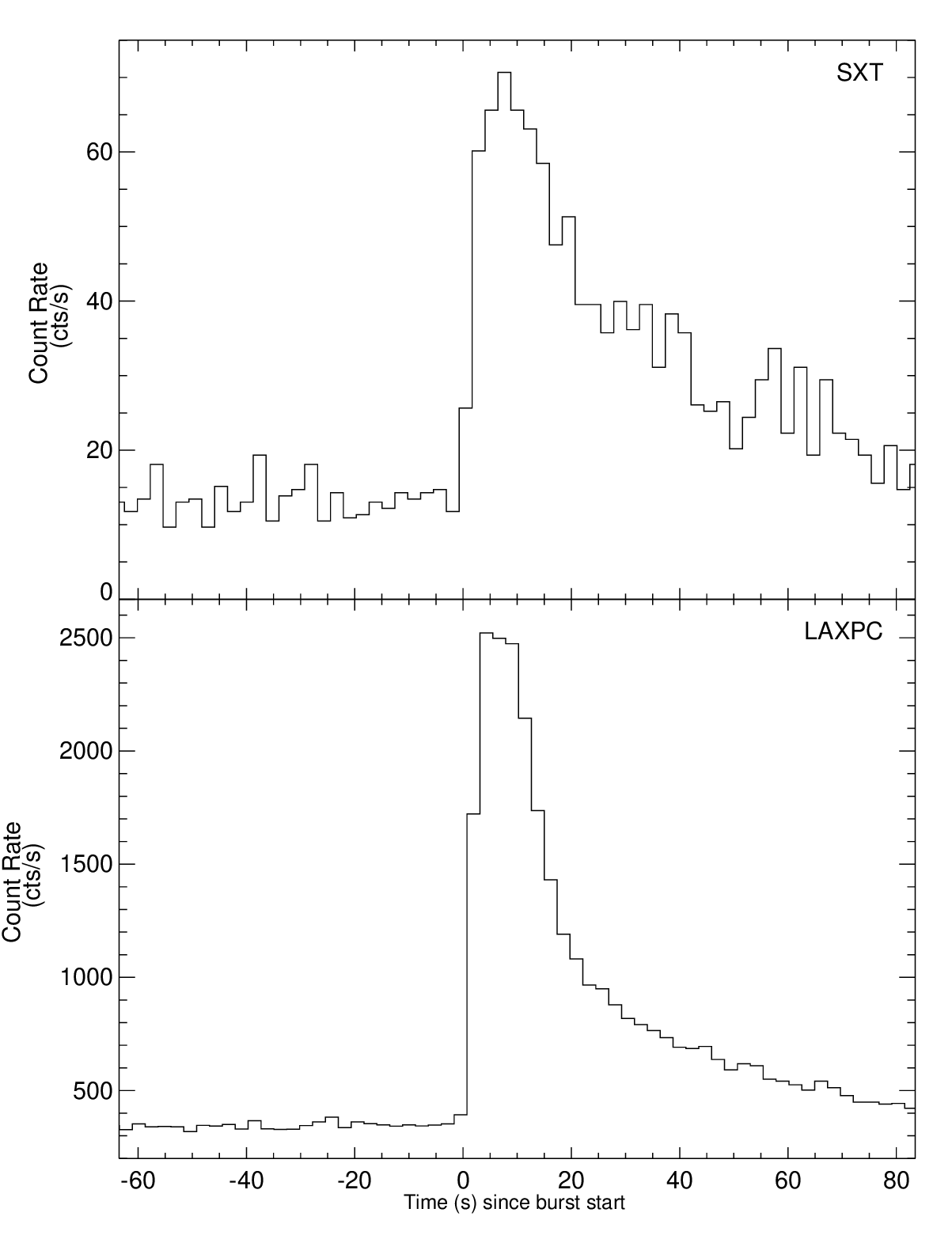}
     \includegraphics[width=0.40\textwidth]{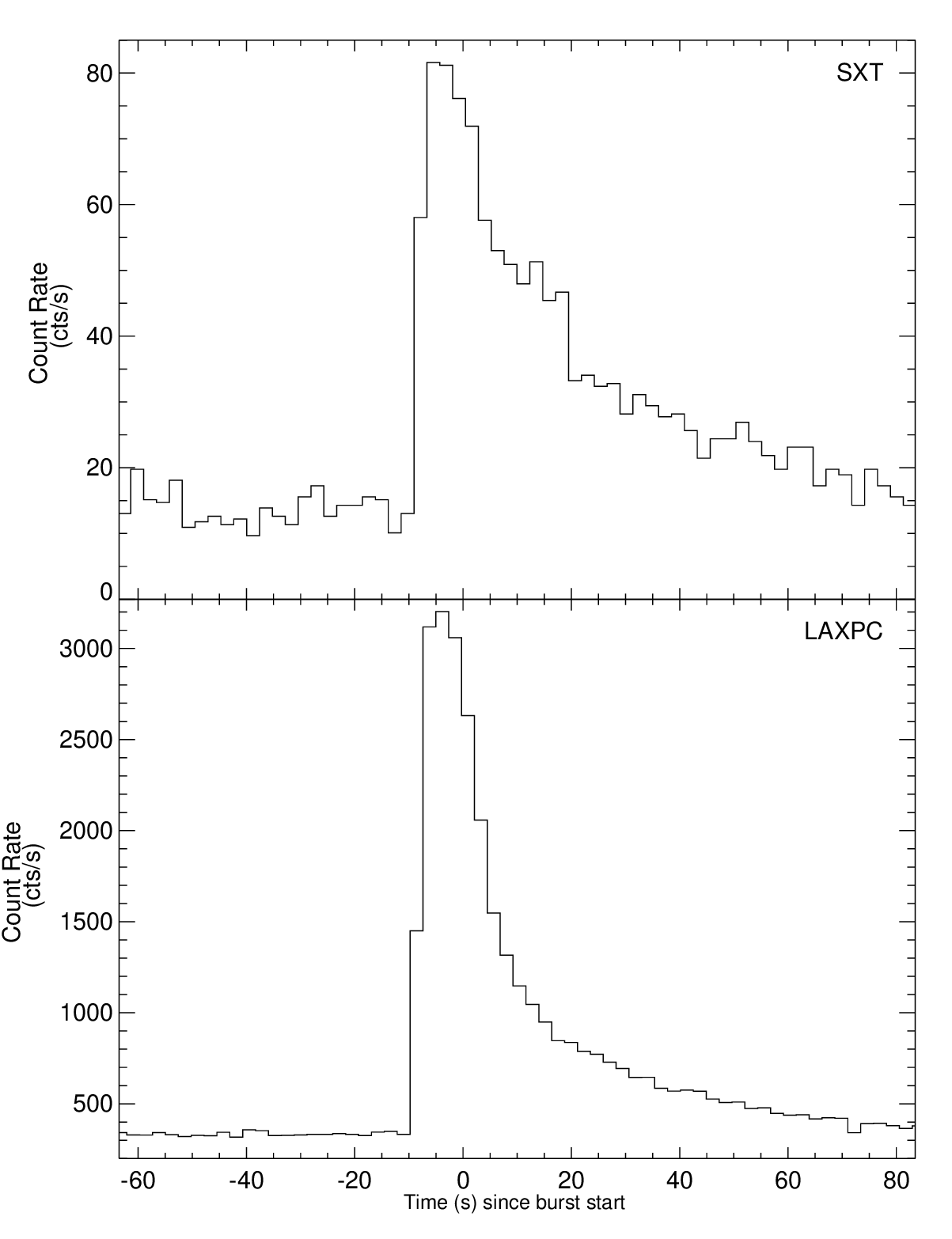}
      \includegraphics[width=0.40\textwidth]{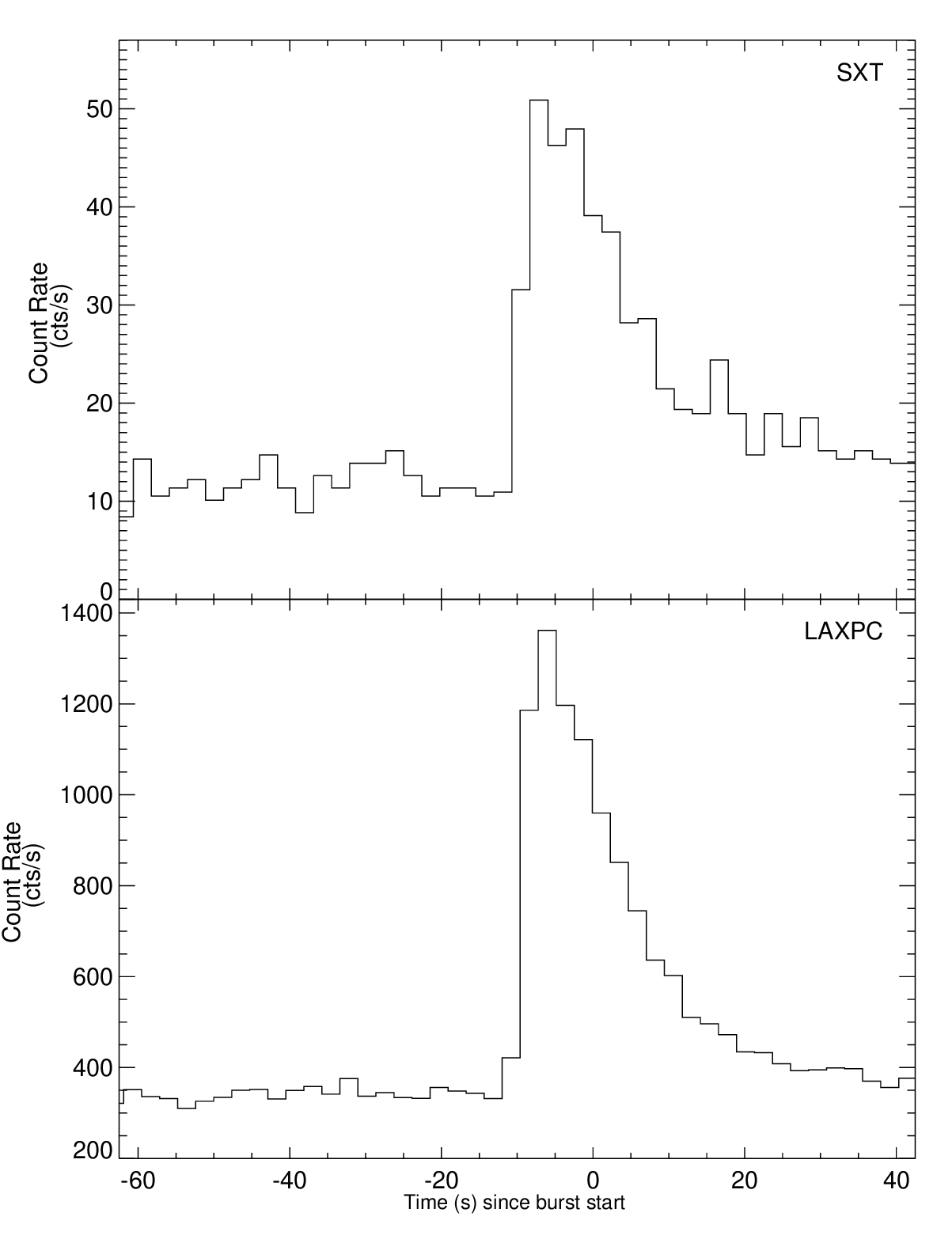}
     \caption{Top left: B5, Top right: B6, Bottom left: B7, Bottom right: B8}
    
\end{figure*}

\addtocounter{figure}{ -1}
\begin{figure*}
    \centering
     \includegraphics[width=0.40\textwidth]{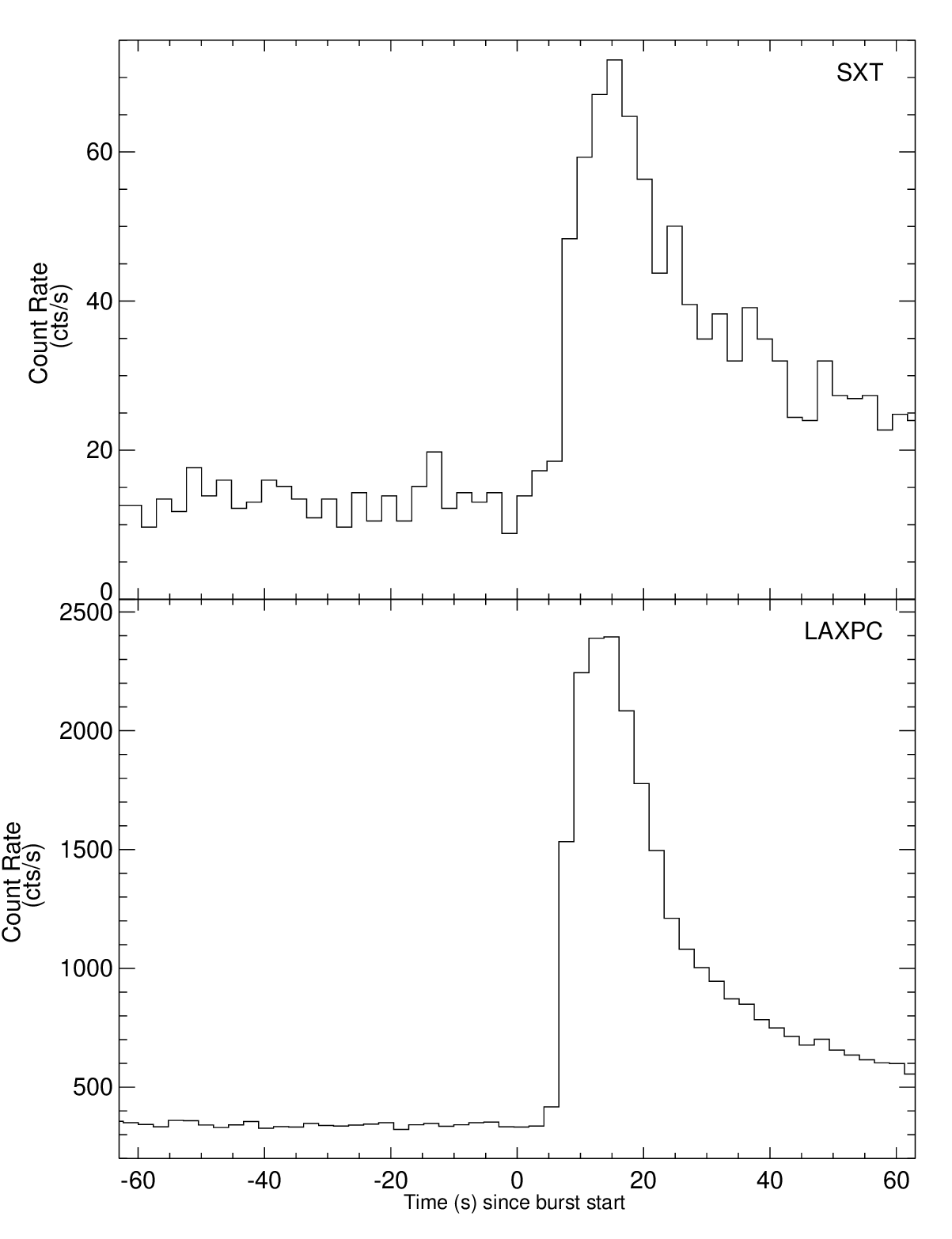}
     \includegraphics[width=0.40\textwidth]{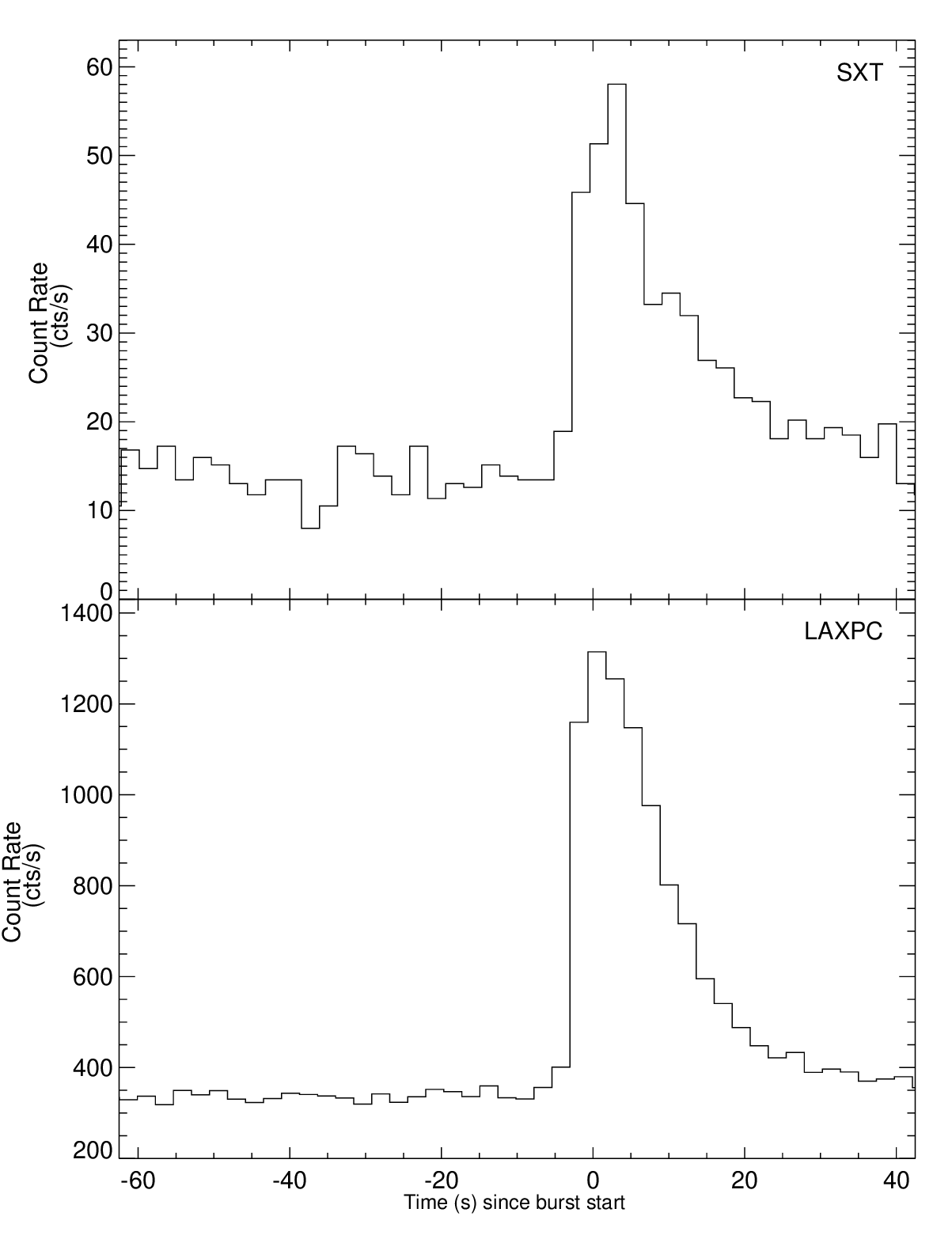}
     \includegraphics[width=0.40\textwidth]{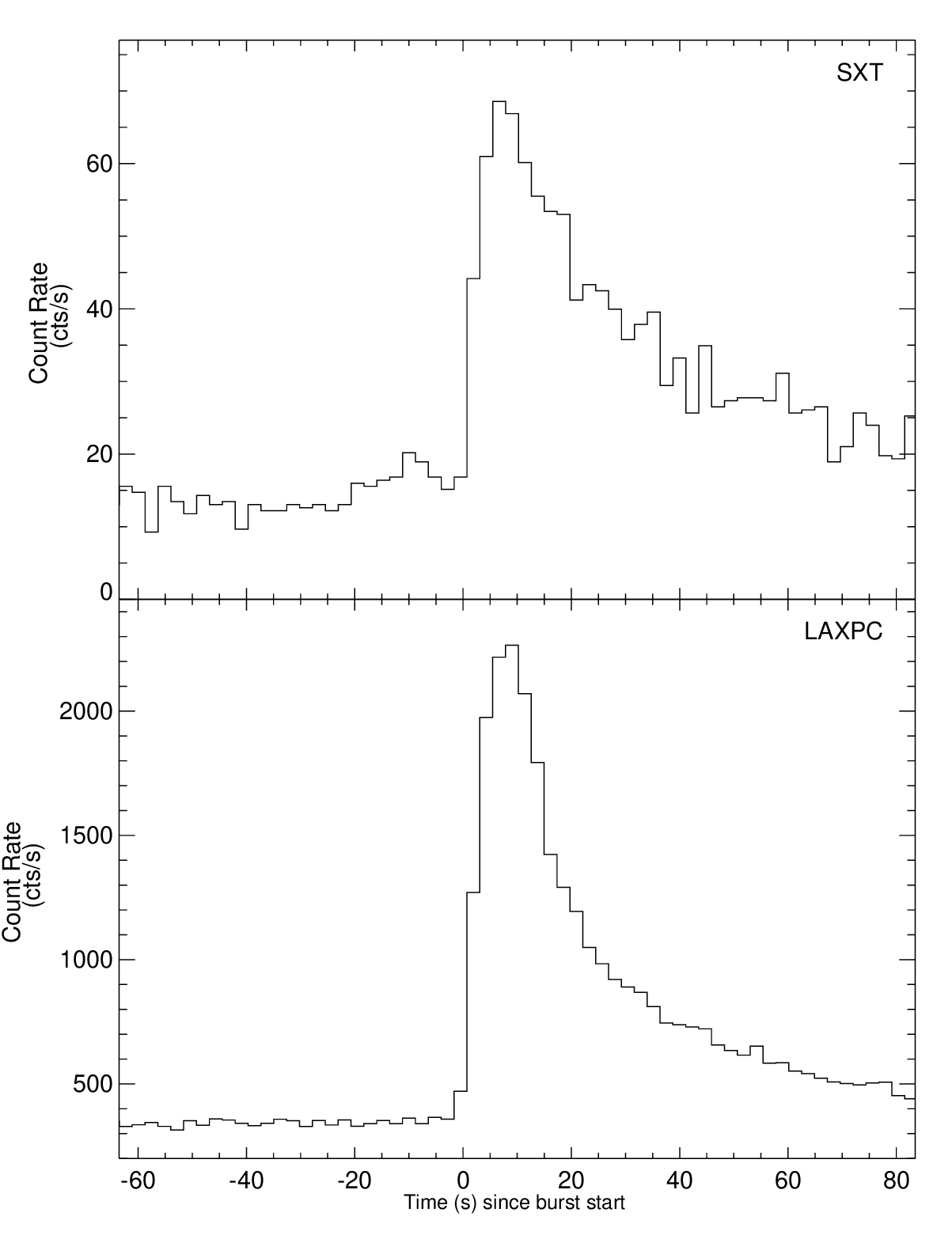}
     \caption{Top left: B9, Top right: B10, Bottom: B11}
    
\end{figure*}

\addtocounter{figure}{ -1}
\begin{figure*}
    \centering
     \includegraphics[width=0.40\textwidth]{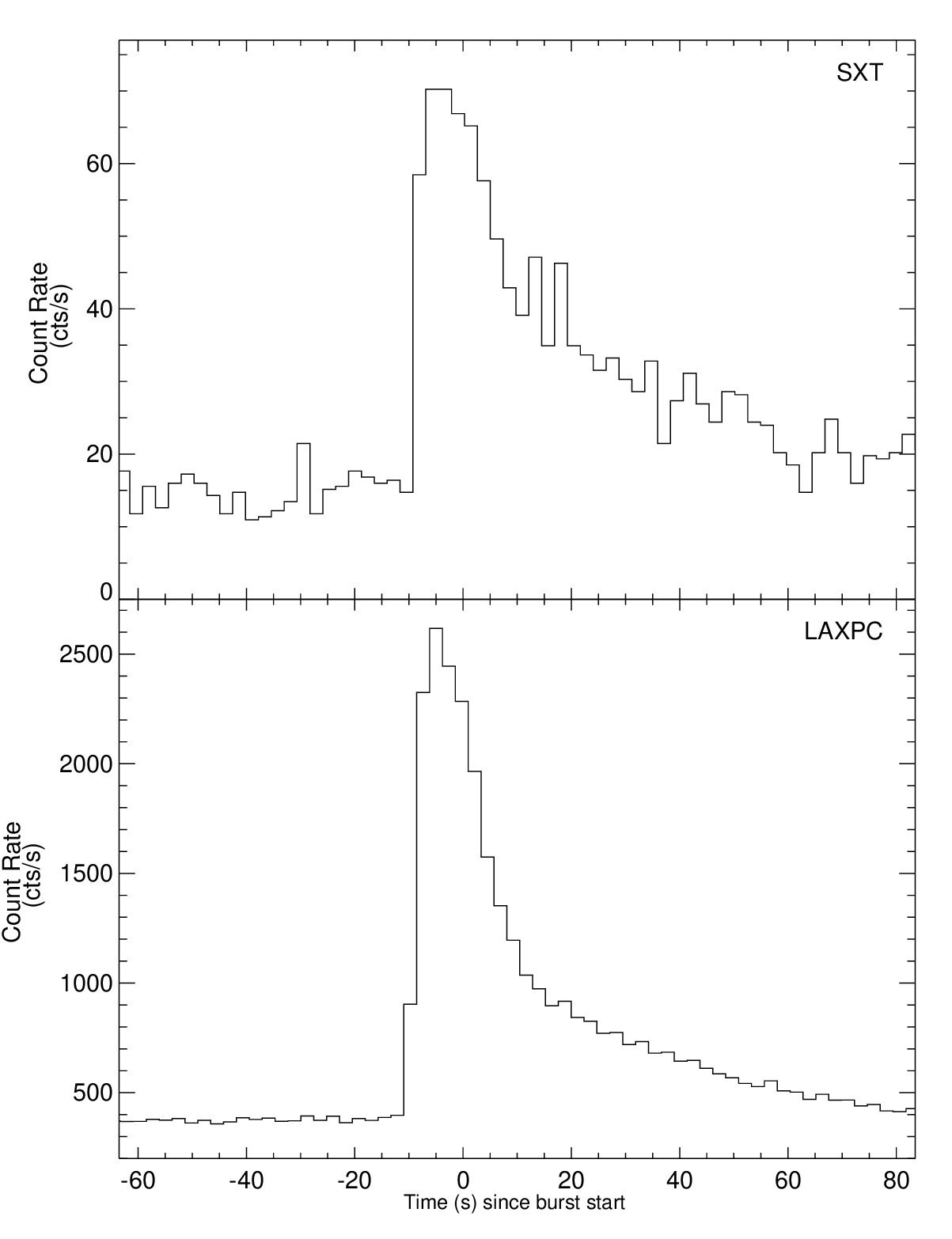}
     \includegraphics[width=0.40\textwidth]{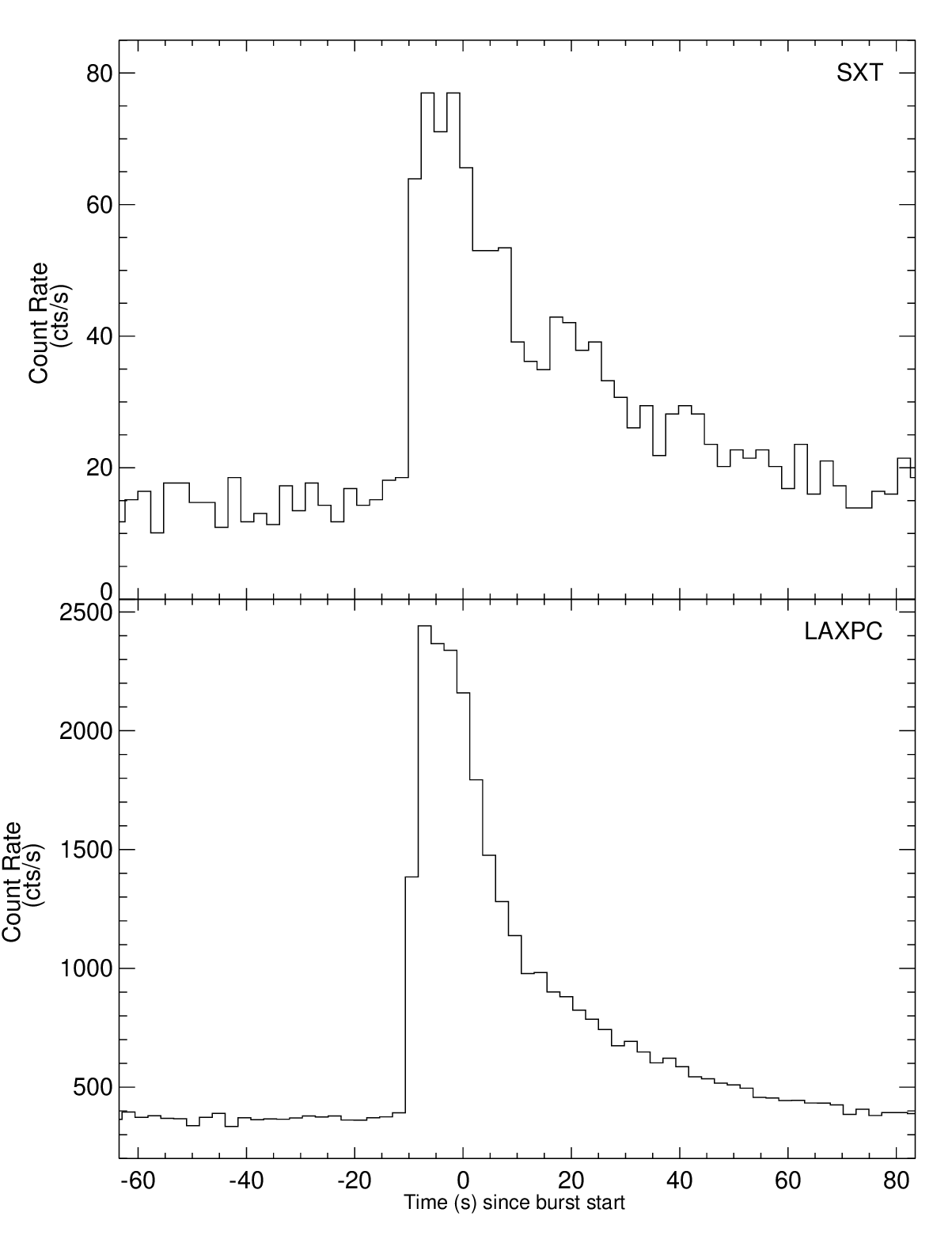}
     \includegraphics[width=0.40\textwidth]{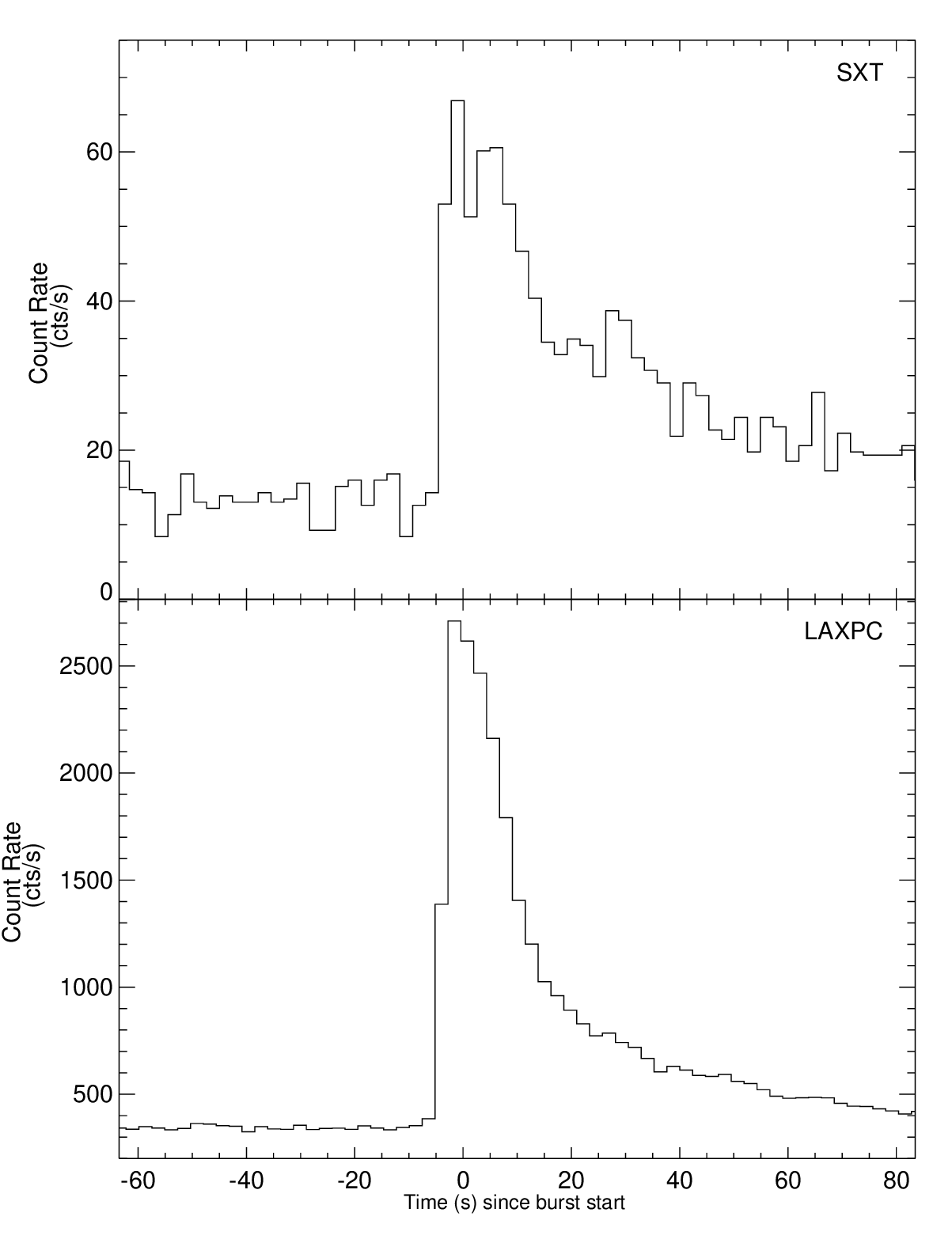}
     \includegraphics[width=0.40\textwidth]{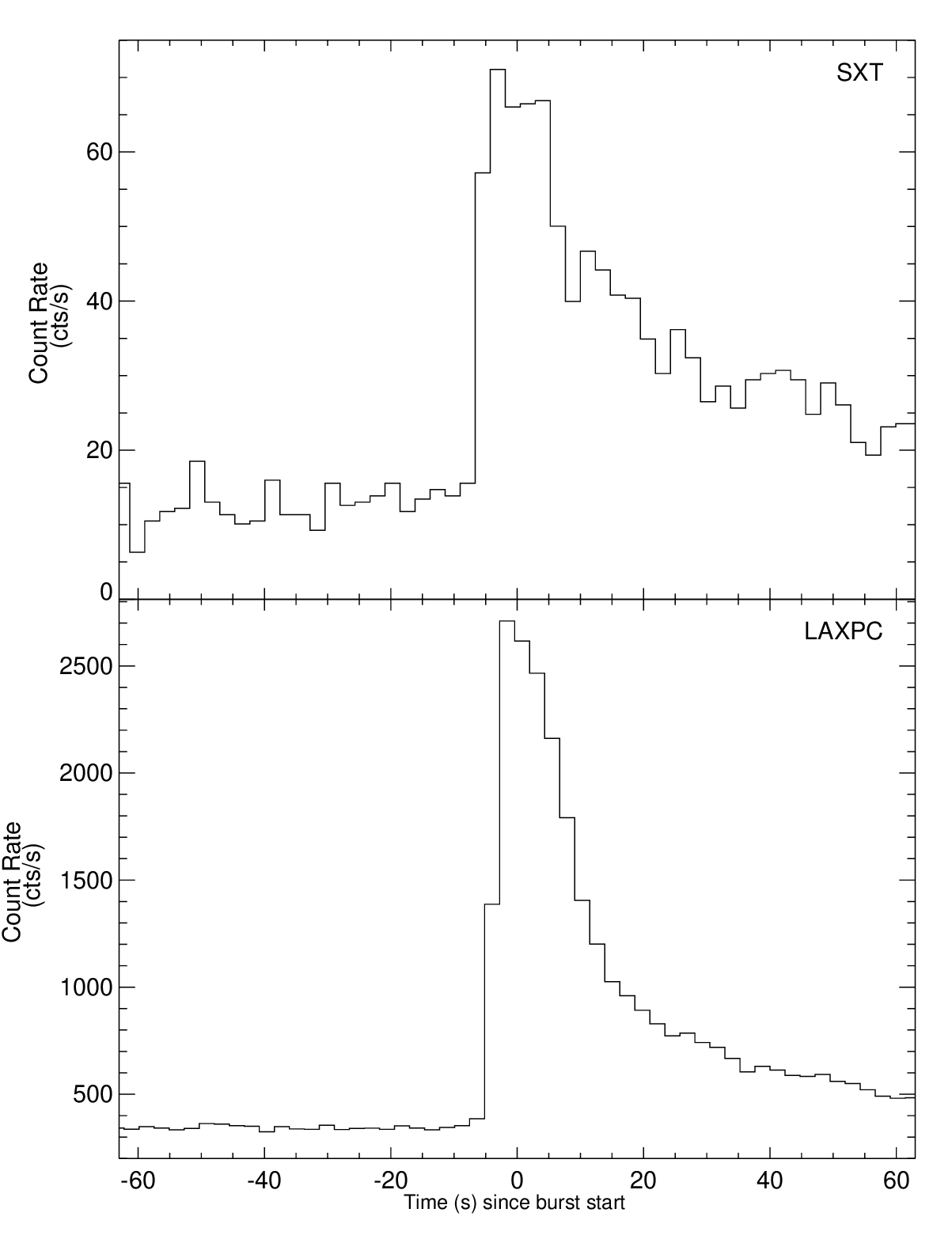}
     \caption{Top left: B12, Top right: B13, Bottom left: B14, Bottom right: B15    }
   
\end{figure*}

\clearpage

\section{Best fitting parameters from time-resolved spectroscopic analysis of all the 15 bursts detected  with single and double blackbody models. }

\begin{figure*}  
        \begin{tabular}{lr}
         \includegraphics[width=0.49\textwidth]{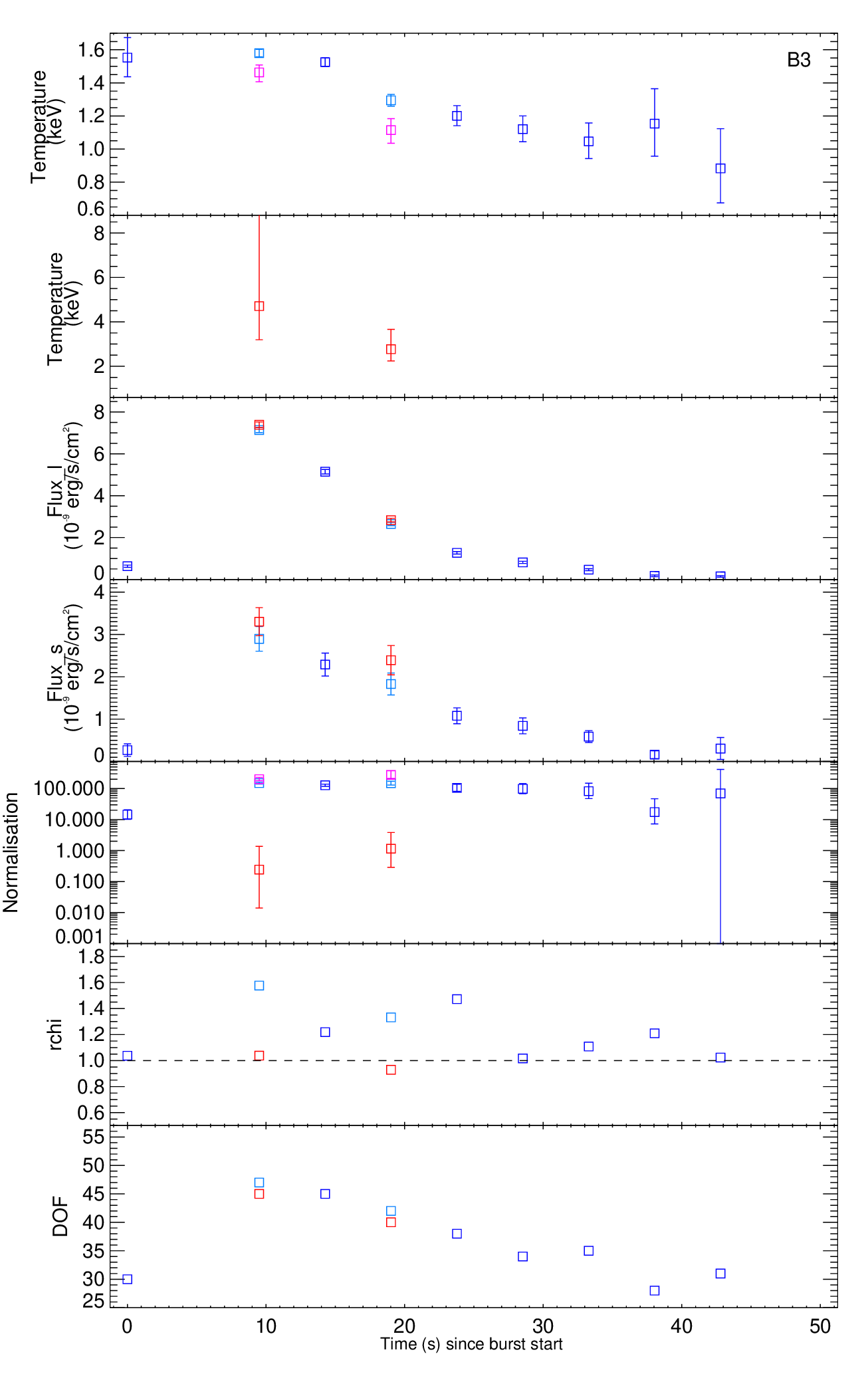} &
         \includegraphics[width=0.49\textwidth]{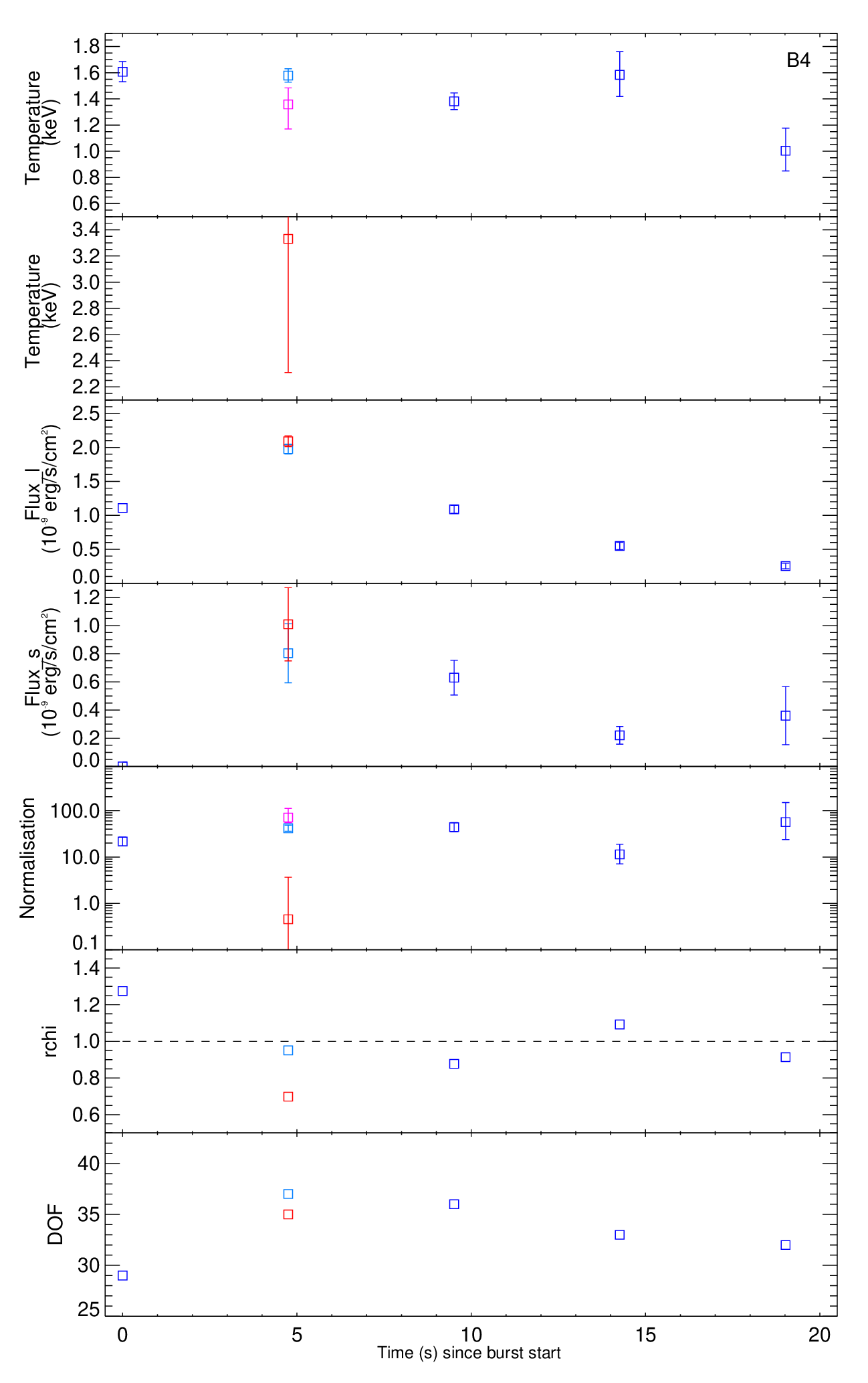}
         \end{tabular}
         \caption{Continued.}
\end{figure*}           
\clearpage  
\addtocounter{figure}{ -1}
\begin{figure*}  
         \begin{tabular}{lr}
         \includegraphics[width=0.49\textwidth]{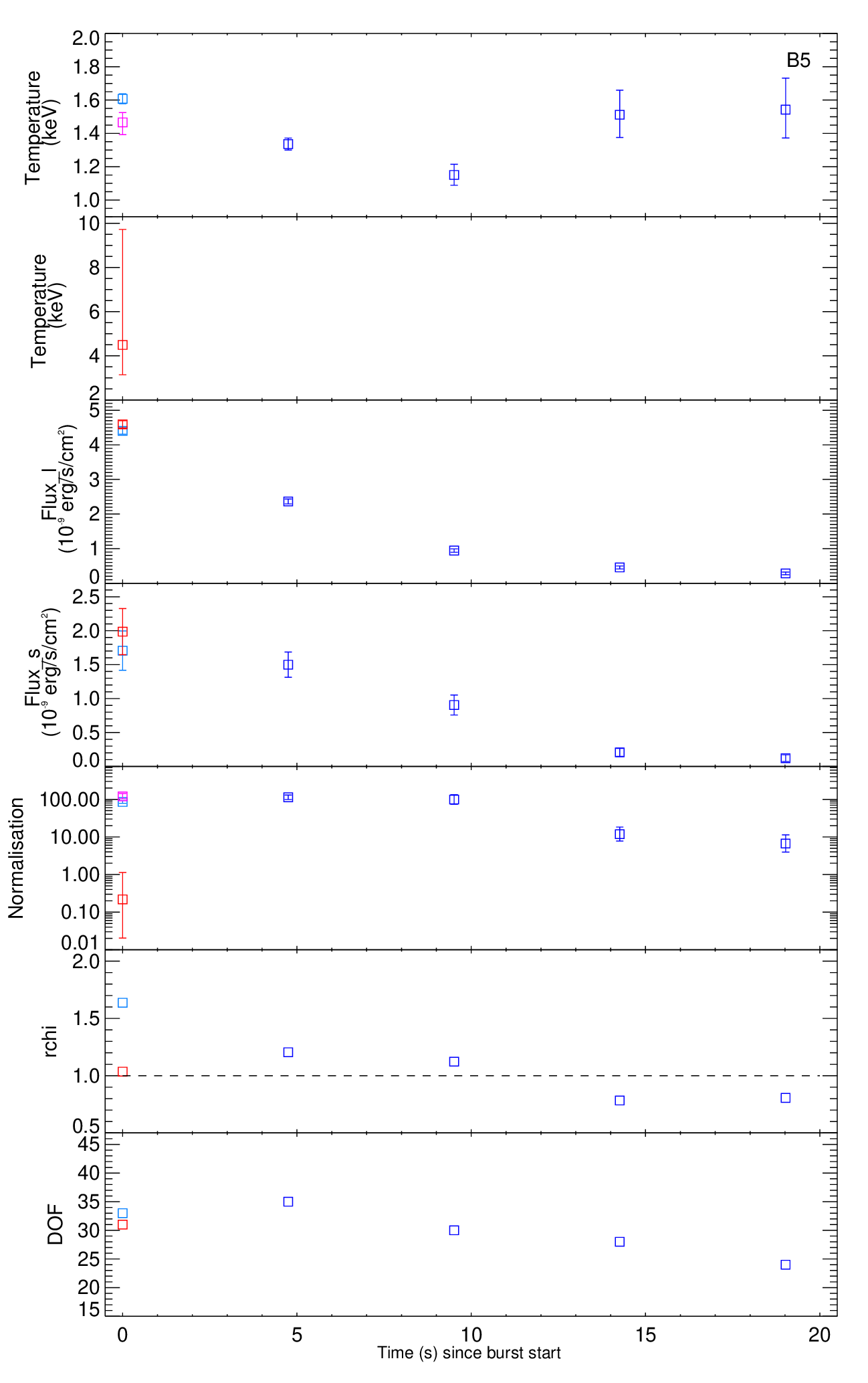} &
         \includegraphics[width=0.49\textwidth]{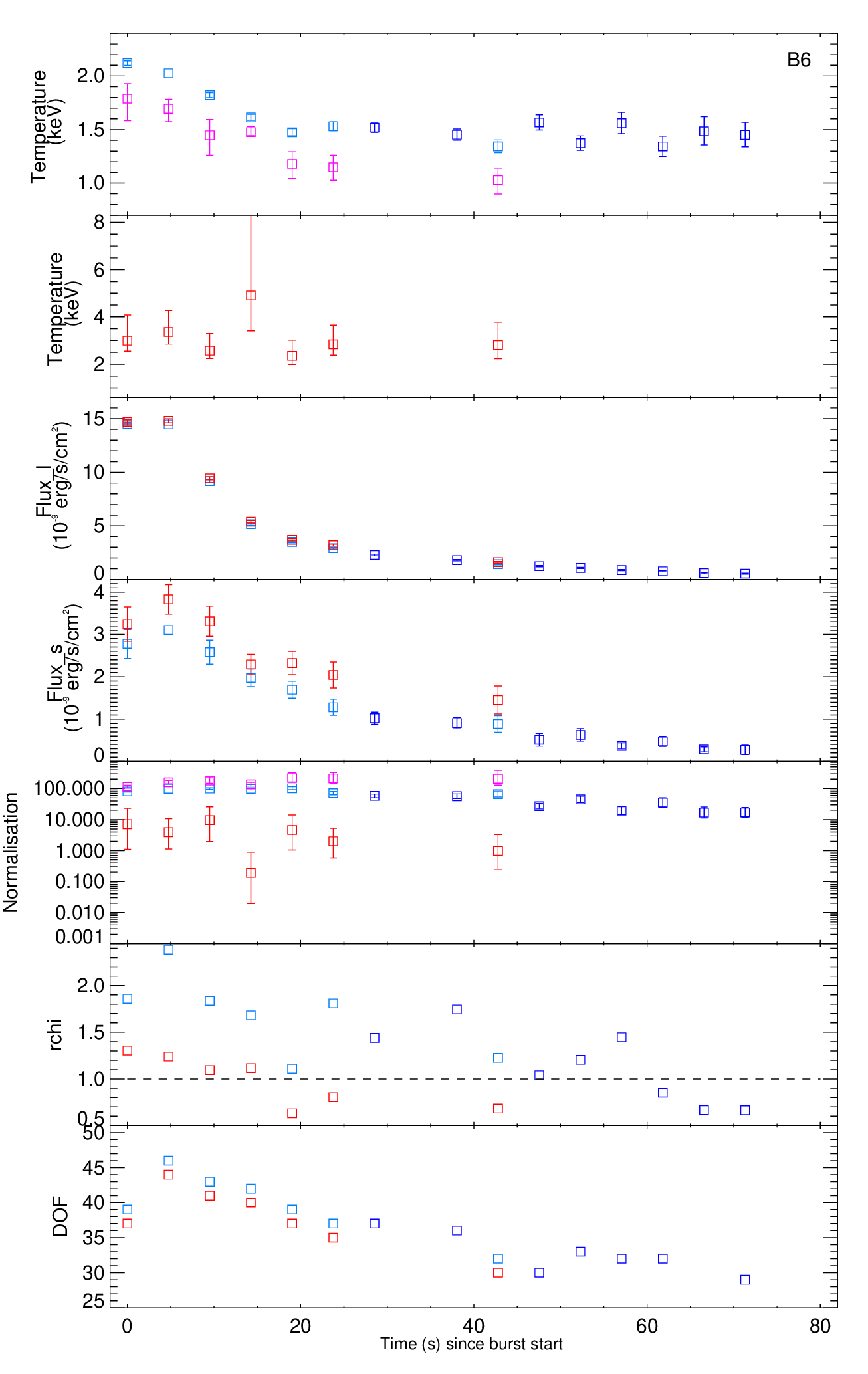}
         \end{tabular}
         \caption{Continued.}
\end{figure*}  
\clearpage
\addtocounter{figure}{ -1}
\begin{figure*}
         \begin{tabular}{lr}
         \includegraphics[width=0.49\textwidth]{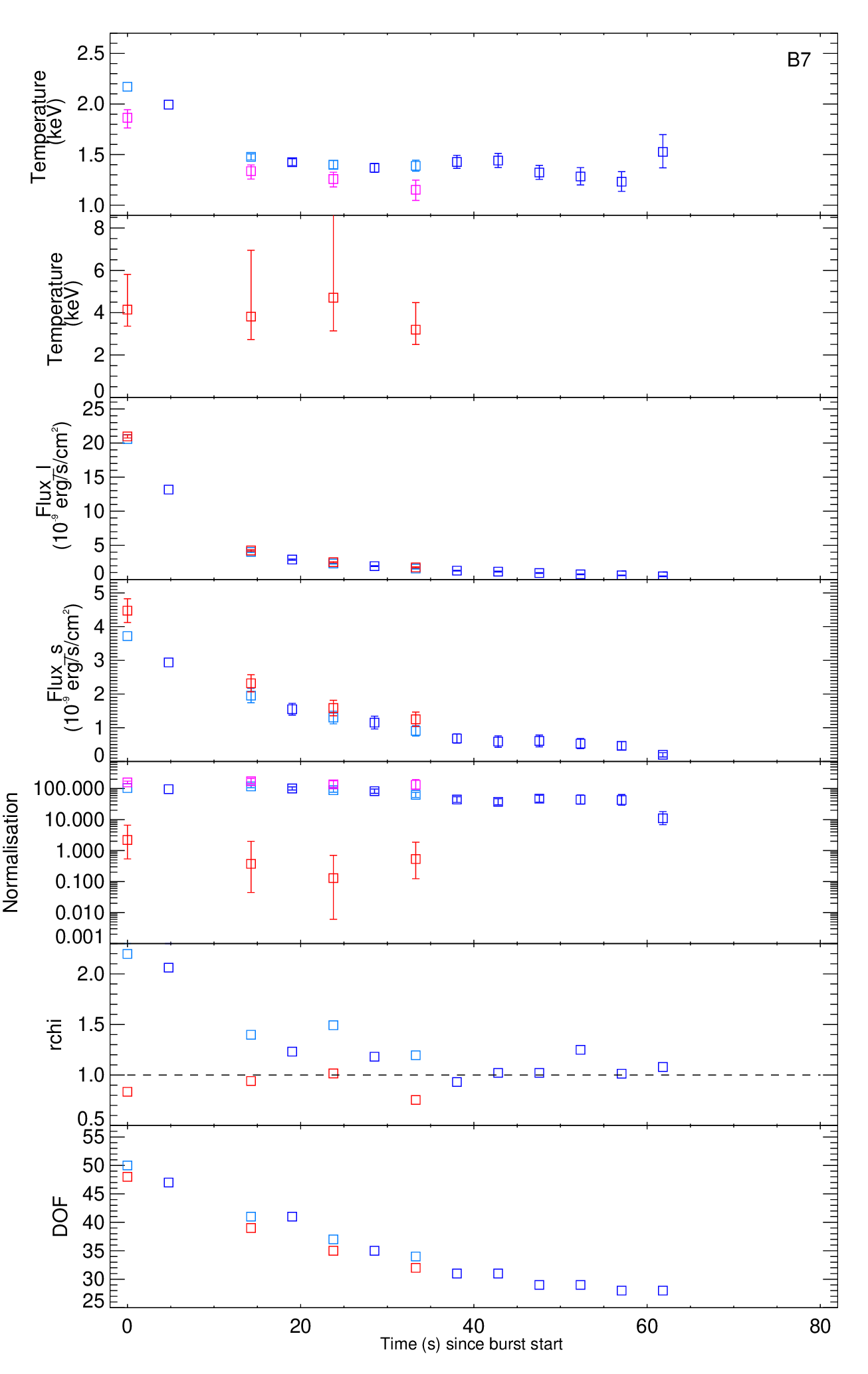}  &
         \includegraphics[width=0.49\textwidth]{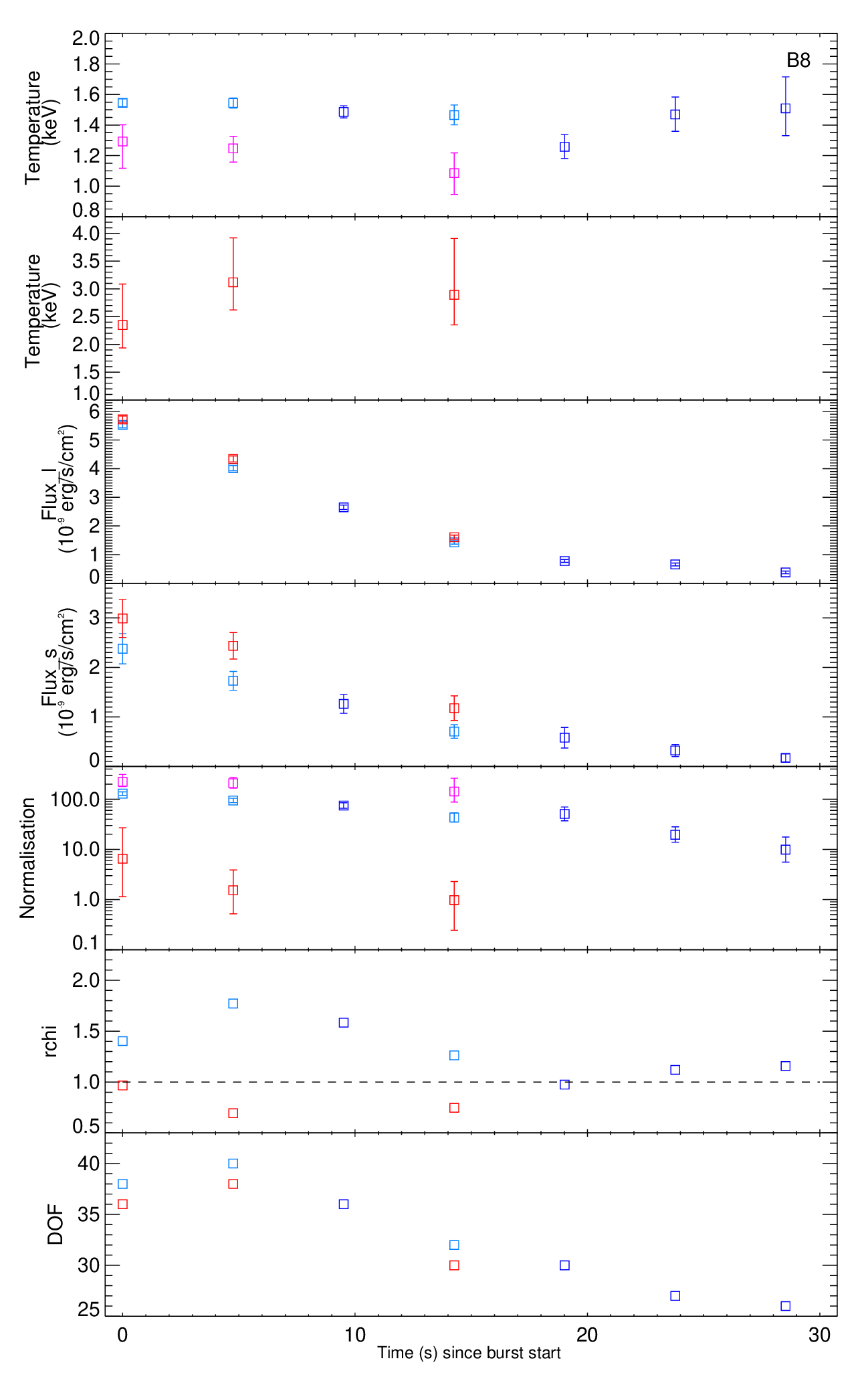}
         \end{tabular}
         \caption{Continued.}
         \label{B7}
\end{figure*}
\clearpage
\addtocounter{figure}{ -1}
\begin{figure*}
         \begin{tabular}{lr}
         \includegraphics[width=0.49\textwidth]{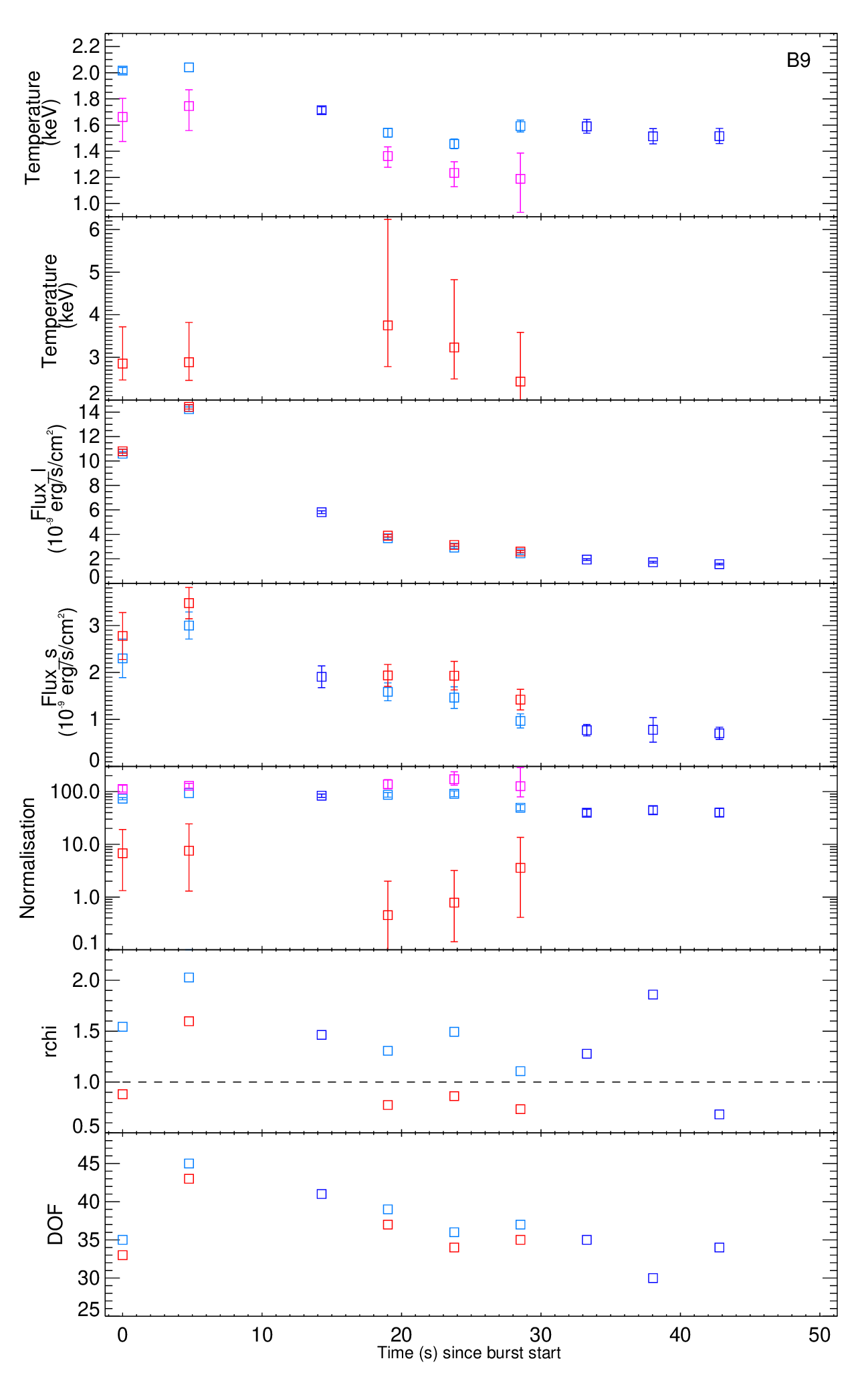} &
         \includegraphics[width=0.49\textwidth]{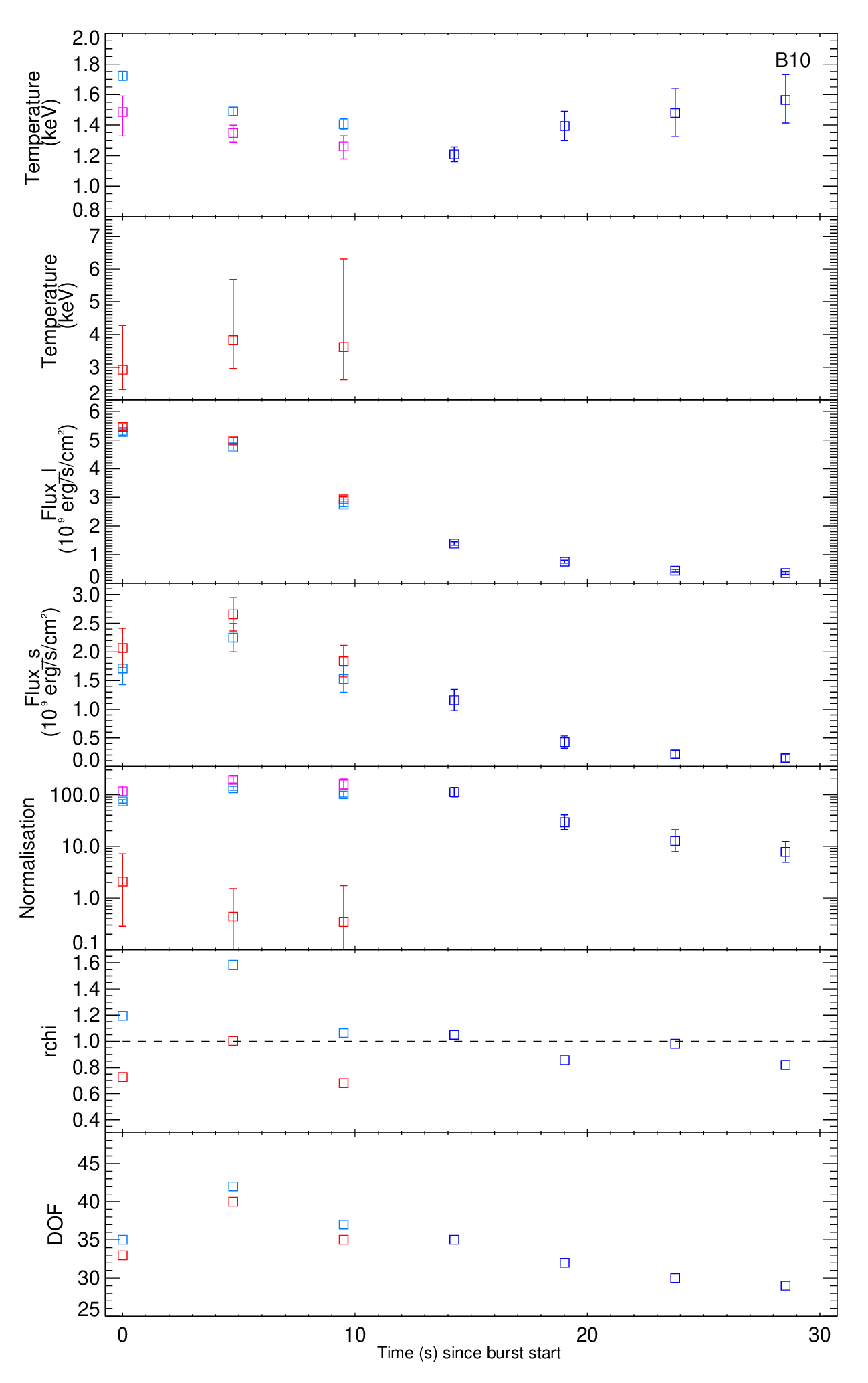}
         \end{tabular}
         \caption{Continued.}
\end{figure*}
\clearpage
\addtocounter{figure}{ -1}
\begin{figure*}
         \begin{tabular}{lr}
         \includegraphics[width=0.49\textwidth]{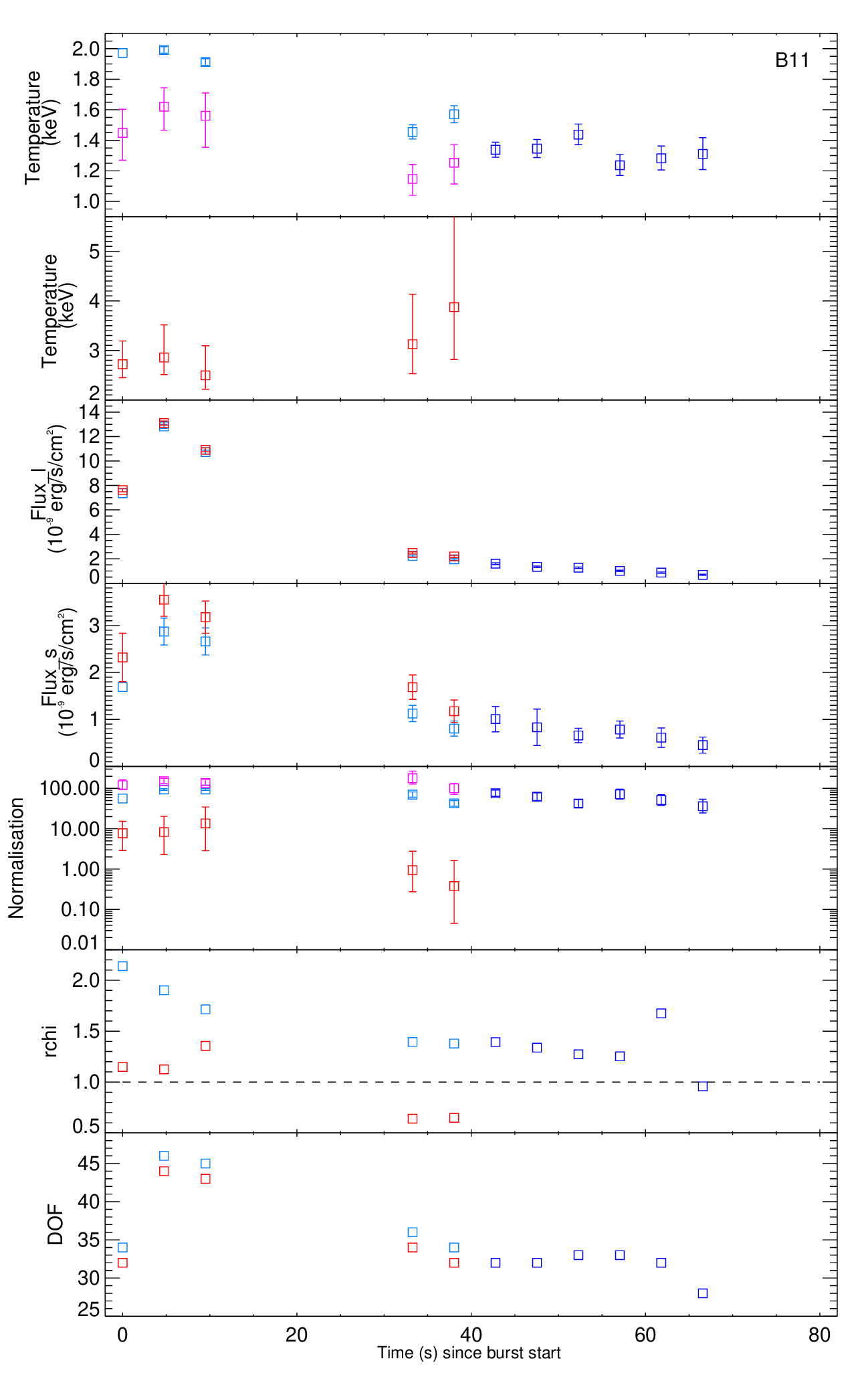} &
         \includegraphics[width=0.49\textwidth]{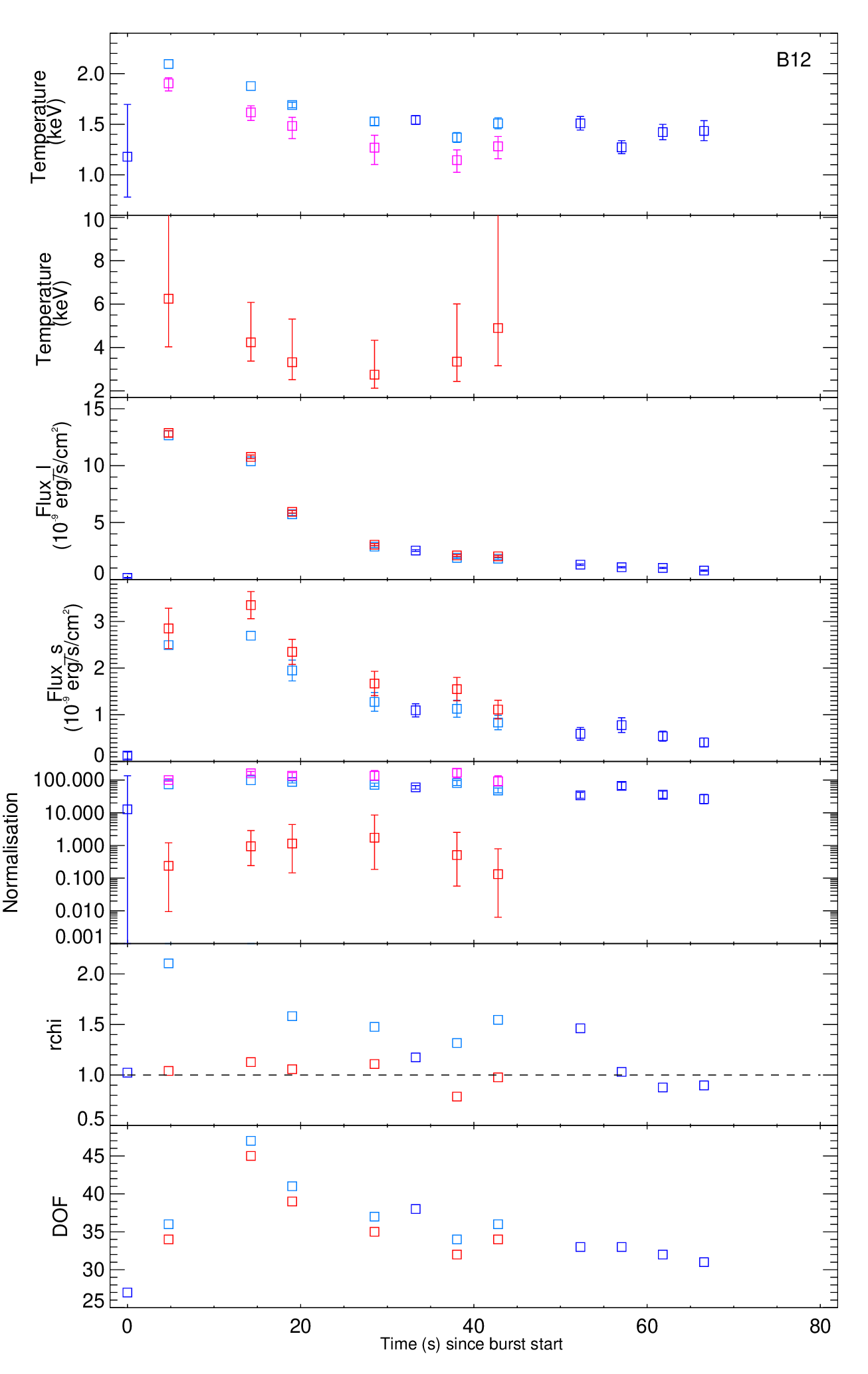}
         \end{tabular}
         \caption{Continued.}
\end{figure*}
\clearpage
\addtocounter{figure}{ -1}
\begin{figure*}
         \begin{tabular}{lr}
         \includegraphics[width=0.49\textwidth]{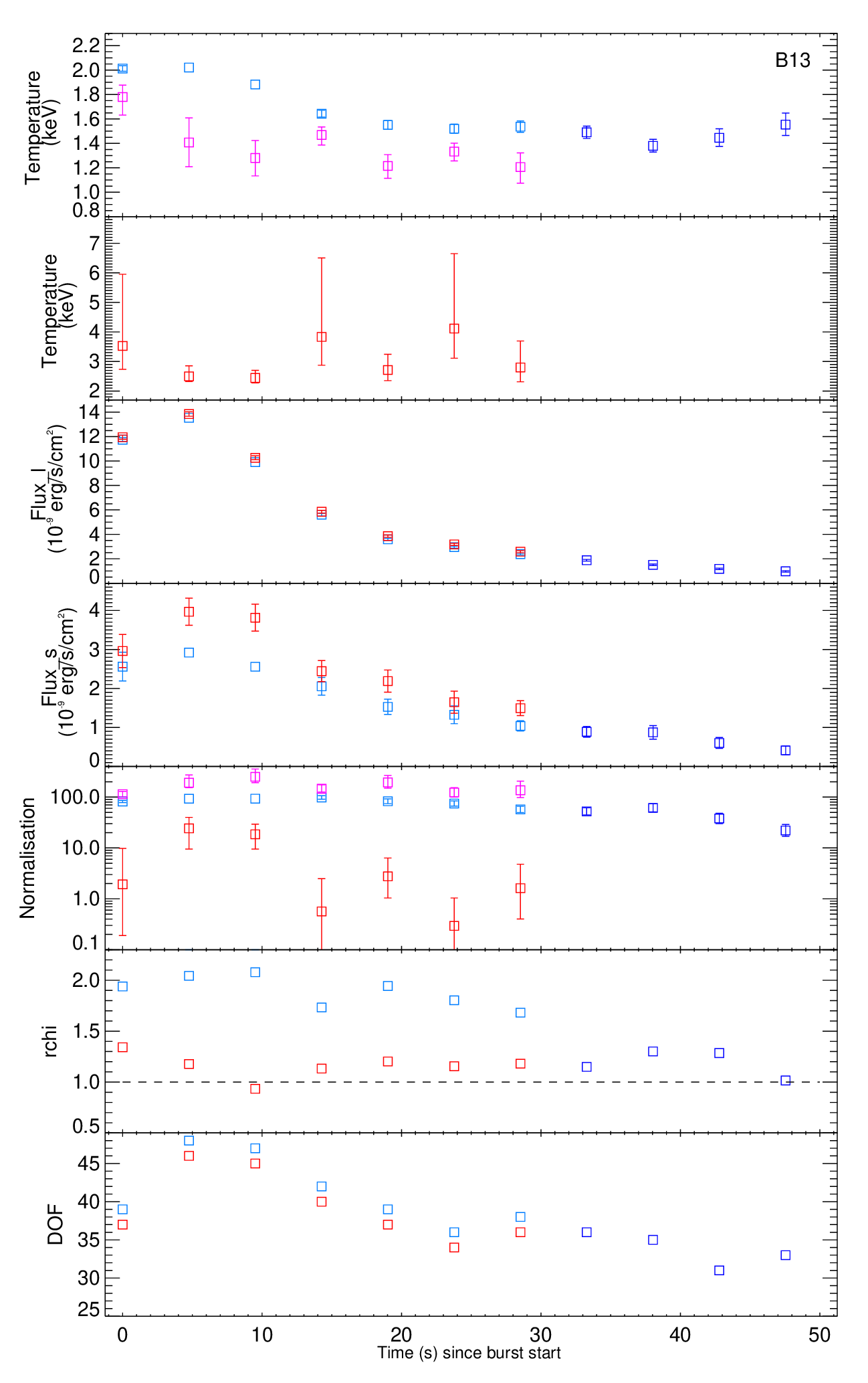} &
         \includegraphics[width=0.49\textwidth]{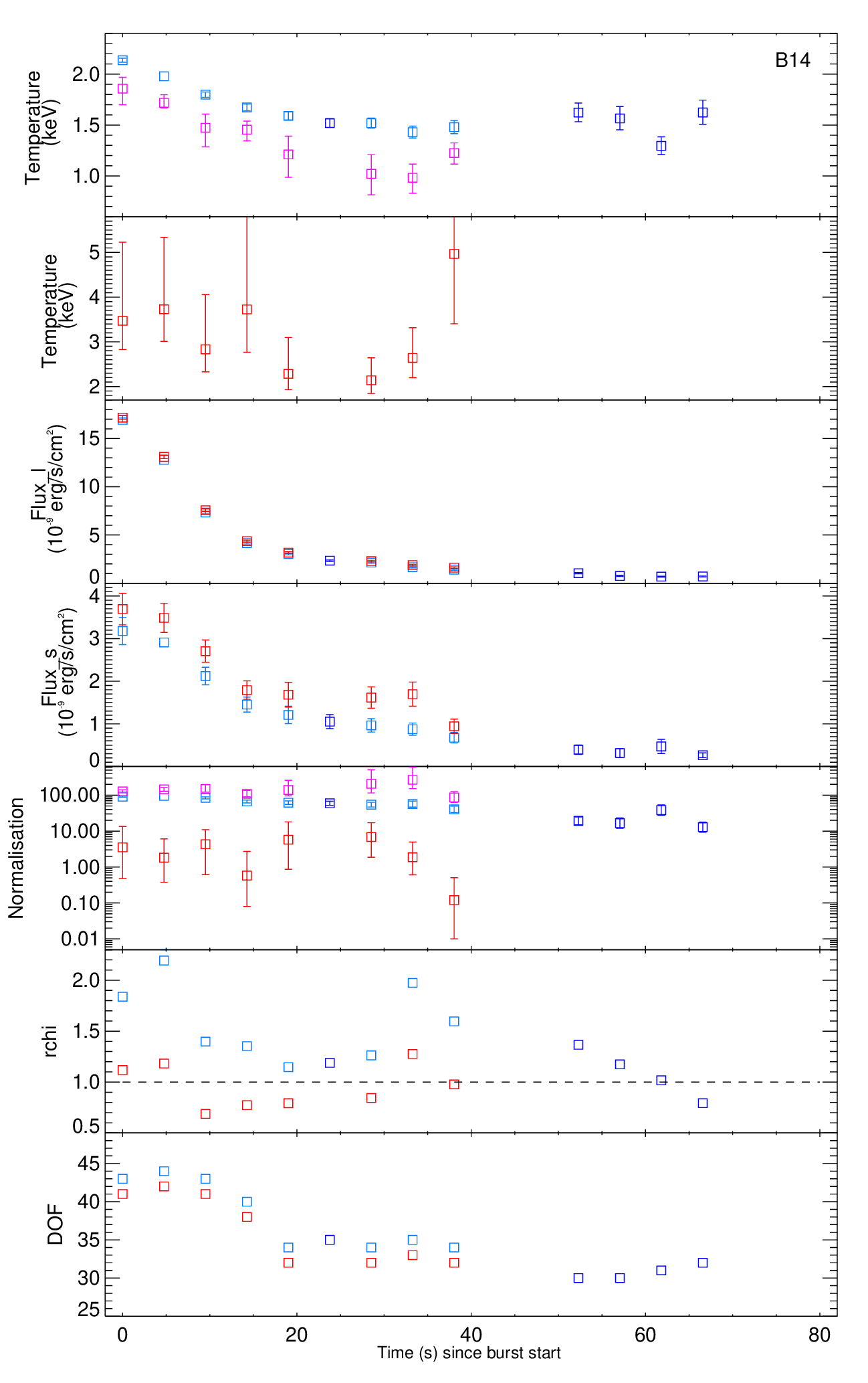}
         \end{tabular}
         \caption{Continued.}
\end{figure*}
\clearpage
\addtocounter{figure}{ -1}
\begin{figure}
         \begin{tabular}{l}
         \includegraphics[width=0.49\textwidth]{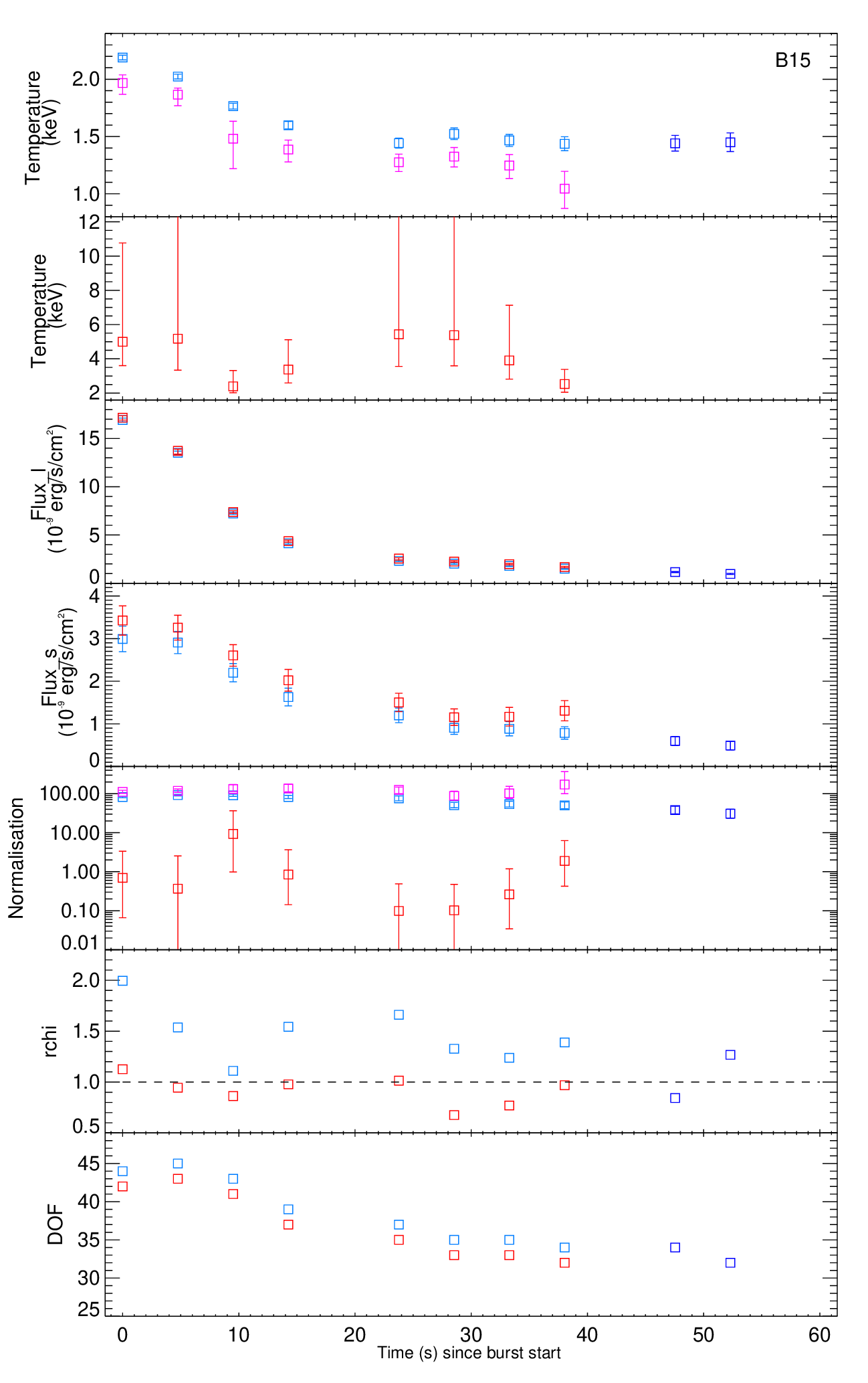} 
         \end{tabular}
         \caption{Continued.}
         
\end{figure}


\bsp	
\label{lastpage}
\end{document}